\documentclass[10pt]{iopart}

\usepackage{graphicx,cite,amssymb}

\def\<{\langle}
\def\>{\rangle}

\newcommand{\PA}{{\it Physica }}
\newcommand{\NAT}{{\it Nature }}
\newcommand{\PRA}{\PR~A }
\newcommand{\PRB}{\PR~B }
\newcommand{\PRE}{\PR~E }
\newcommand{\bc}{BC}

\newcommand{\upd}{{\ensuremath{\textrm{d}}}}
\usepackage{upgreek}

\newcommand{\pp}{\ensuremath{{(+,+)}}}

\renewcommand*{\vec}[1]{\mathbf{#1}}
\newcommand{\eeref}[1]{equation \eref{#1}}

\newcommand{\Tabletwo}[1]{\begin{table*}
  \caption{#1}
  \begin{indented}
  \lineup
  \item[]
\begin{center}
\begin{tabular}{@{}l*{15}{l}}}
\def\endTabletwo{\end{tabular}\end{center}\end{indented}\end{table*}}

\begin{document}
\title[Line contribution to the critical Casimir force]{Line contribution to the critical Casimir force between a homogeneous and a chemically stepped surface}
\author{Francesco Parisen Toldin$^1$, Matthias Tr\"ondle$^{2,3}$ and S Dietrich$^{2,3}$}
\address{$^1$Institut f\"ur Theoretische Physik und Astrophysik, Universit\"at W\"urzburg, Am Hubland, D-97074 W\"urzburg, Germany}
\ead{francesco.parisentoldin@physik.uni-wuerzburg.de}
\address{$^2$Max-Planck-Institut f\"ur Intelligente Systeme, Heisenbergstr.~3, D-70569 Stuttgart, Germany}
\address{$^3$IV. Institut f\"ur Theoretische Physik, Universit\"at Stuttgart, Pfaffenwaldring 57, D-70569 Stuttgart, Germany}
\ead{troendle@is.mpg.de}
\ead{dietrich@is.mpg.de}

\begin{abstract}
Recent experimental realizations of the critical Casimir effect have been implemented by monitoring colloidal particles immersed in a binary liquid mixture near demixing and exposed to a chemically structured substrate. In particular, critical Casimir forces have been measured for surfaces consisting of stripes with periodically alternating adsorption preferences, forming chemical steps between them. Motivated by these experiments, we analyze the contribution of such chemical steps to the critical Casimir force for the film geometry and within the Ising universality class. By means of Monte Carlo simulations, mean-field theory, and finite-size scaling analysis we determine the universal scaling function associated with the contribution to the critical Casimir force due to individual, isolated chemical steps facing a surface with homogeneous adsorption preference or with Dirichlet boundary condition. In line with previous findings, these results allow one to compute the critical Casimir force for the film geometry and in the presence of arbitrarily shaped, but wide stripes. In this latter limit the force decomposes into a sum of the contributions due to the two homogeneous parts of the surface and due to the chemical steps between the stripes. We assess this decomposition by comparing the resulting sum with actual simulation data for the critical Casimir force in the presence of a chemically striped substrate.
\end{abstract}

\pacs{05.70.Jk, 64.60.an, 68.15.+e, 05.50.+q, 05.10.Ln}

\noindent{\it Keywords\/}: Critical Casimir force, Critical phenomena, Monte Carlo simulations, Mean-field theory, Finite-size scaling, Liquid thin films

\ioptwocol
\section{Introduction}
\label{sec:intro}
In a fluid, the spatial extent of fluctuations, which is given by the correlation length, can become macroscopically large if the fluid approaches a critical point. Under such thermodynamic conditions, the confinement of thermal fluctuations results in an effective force between the confining surfaces. The occurrence of this particular fluctuation-induced force, first predicted by Fisher and de~Gennes \cite{FG-78}, is known as the critical Casimir effect which is the analogue of the Casimir effect in quantum electrodynamics \cite{Casimir-48}. Reference \cite{Gambassi-09} provides a recent review which illustrates analogies as well as differences between these two effects. For reviews of the critical Casimir effect see also \cite{Krech-99,Krech-94,BTD-00} and the updated reference list in \cite{PTD-10}.

The critical Casimir force is determined by the bulk and surface universality classes (UC) \cite{Binder-83,Diehl-86} of the confined system. It is characterized by a universal scaling function, which is independent of microscopic details of the system and depends only on a few global and general properties, such as the spatial dimension $d$, the number of components of the order parameter, the shape of the confinement, and the type of boundary conditions ({\bc}) at the confining boundaries of the system \cite{Krech-94,Krech-99,BTD-00}.

The first experimental evidence of the critical Casimir force has been obtained by studying wetting films of fluids close to a critical end point \cite{NI-85,KD-92b}. In this context, $^4$He wetting films close to the onset of superfluidity \cite{he4} and wetting films of classical \cite{cwetting} and quantum \cite{qwetting} binary liquid mixtures have been studied experimentally. More recently direct measurements of the critical Casimir force have been reported \cite{HHGDB-08,GMHNHBD-09,SZHHB-08,NHB-09,TZGVHBD-11,NDHCNVB-11,ZAC-11} by monitoring the Brownian motion of individual colloidal particles immersed into a binary liquid mixture close to its critical demixing point and exposed to a planar wall. The critical Casimir effect has also been studied experimentally via its influence on aggregation phenomena \cite{BOSGWS-09,VAWPMSW-12,NFHWS-13,PMVWMSW-14}.

Early theoretical studies of the  critical Casimir force have used, to a large extent, field-theoretical methods (see, e.g., \cite{PTD-10} for a list of references). Exact results are available for the Ising UC in two dimensions \cite{Cardy-87,ZRSA-10,AM-09,AM-13} and in three dimensions for the spherical model \cite{CDT-97,Dantchev-98,DK-04,CD-04,DDG-05,DG-08} and in the large$-N$ limit \cite{DGHHRS-12,DBR-14}. In three dimensions, only recently their quantitatively reliable computation has been obtained by means of Monte Carlo (MC) simulations. Early numerical simulations for the critical Casimir force have been employed in \cite{Krech-97} for the film geometry with laterally homogeneous {\bc}. More recently, by using MC simulations the critical Casimir force has been computed for the $XY$ UC \cite{DK-04,Hucht-07,VGMD-07,VGMD-08,Hasenbusch-09b,Hasenbusch-09c,Hasenbusch-09d}, which describes the critical properties of the superfluid phase transition in $^4$He, as well as the Ising UC \cite{DK-04,VGMD-07,VGMD-08,Hasenbusch-10c,PTD-10,HGS-11,Hasenbusch-11,VMD-11,Hasenbusch-12,Hasenbusch-12b,VED-13,PTTD-13,PT-13,VD-13,Vasilyev-14,HH-14} which describes, {\it inter alia}, the critical behavior of a binary liquid mixture close to its demixing phase transition.

For the latter system, the involved surfaces typically prefer to adsorb one of the two species of the mixture, leading to symmetry-breaking {\bc} (denoted as $(+)$ or $(-)$ {\bc}) acting on the order parameter which is the deviation of the concentration of one of the two species from its value at the critical point. Not only the shape of the universal scaling function of the critical Casimir force is determined by the {\bc}, but also the sign of the force depends on the combination of the {\bc}. In the case of laterally homogeneous adsorption preferences of the confining surfaces, the force is attractive if the adsorption preferences are the same, i.e., for so-called $(+, +)$ {\bc}, whereas it is repulsive for opposite adsorption preferences, i.e., for so-called $(+,-)$ {\bc}. This result has been first predicted by mean field theory in \cite{Krech-97}, later confirmed by MC simulations \cite{VGMD-07,VGMD-08,Hasenbusch-10c}, and experimentally realized in \cite{cwetting,HHGDB-08,GMHNHBD-09}.

Experiments with binary liquid mixtures have been used to study critical Casimir forces acting on colloidal particles close to substrates exhibiting inhomogeneous adsorption preferences \cite{TZGVHBD-11,SZHHB-08,NHB-09}, in particular, for the case of a chemically structured substrate \cite{SZHHB-08} which, for both components of the solvent, creates a laterally varying adsorption preference. Such kind of systems have recently attracted particular interest \cite{TZGVHBD-11,GD-11}. Theoretical investigations have have been focused on the film geometry within mean-field theory \cite{SSD-06}, within Gaussian approximation \cite{KPHSD-04,ZRK-04}, and recently by MC simulations \cite{PTD-10,PTTD-13}. Within the Derjaguin approximation the critical Casimir force in the presence of a chemically patterned substrate has also been studied in the case of a sphere close to a planar wall \cite{TKGHD-09,TKGHD-10}, and in the case of a cylindrical colloid \cite{LLTHD-14}.

Motivated by the experimental results for chemically structured substrate, in a previous paper by two of the authors \cite{PTD-10} we have computed the critical Casimir force for the film geometry in the Ising UC. We have considered a laterally homogeneous adsorption preference for the upper confining surface, whereas the lower surface is divided into two halves, with opposing adsorption preferences and a straight chemical step between them. For this system we employed laterally periodic {\bc}, which give rise to an additional, second chemical step at the lateral boundaries. This geometry is shown in \fref{cs}. In \cite{PTD-10} we have shown that the critical Casimir force decomposes into a sum of the force due to the two homogeneous halves, and a contribution which is solely due to the two individual chemical steps. This result allows one to compute the critical Casimir force also in the case of a chemically striped substrate, consisting of stripes of alternating adsorption preference, provided that the width of the stripes is sufficiently large relative to the thickness of the film. In \cite{PTTD-13} we have determined the critical Casimir force in the actual presence of a chemically striped surface, verifying the validity of the aforementioned decomposition for wide stripes. Here we study the critical Casimir force for the film geometry with the lower surface displaying a single chemical step, while the opposing surface carries {\it Dirichlet} {\bc}; this geometry is illustrated in \fref{csopen}. By means of MC simulations and mean-field theory we extract the chemical step contribution to the critical Casimir force for such {\bc}. We also provide improved results for the case of a chemical step opposing a surface with laterally homogeneous {\it adsorption preference}, which has been considered in \cite{PTD-10}. These results allow us to compute the critical Casimir force in the presence of arbitrarily shaped chemical stripes of large widths. We test this approximation by comparing the resulting force with the corresponding MC results in \cite{PTTD-13}.

This paper is organized as follows. In \sref{sec:fss} we recall the finite-size scaling behavior which allows one to define the critical Casimir force, focusing in particular on the contributions of the chemical steps for the {\bc} shown in figures \ref{cs} and \ref{csopen}. In \sref{sec:mc} we present our MC results, and in \sref{sec:mft} the corresponding mean-field  scaling functions are determined. We summarize our main findings in \sref{sec:summary}.

\section {Finite-size scaling}
\label{sec:fss}
\subsection{General properties}
\label{sec:fss:general}
Here we study such systems in the three-dimensional film geometry of size $L\times L_\parallel\times L_\parallel$ which in the thermodynamic limit exhibit second-order phase transitions in the Ising universality class. We impose periodic {\bc} in the two lateral directions, and various, in parts inhomogeneous {\bc} in the remaining perpendicular direction, to be discussed below. In this section we summarize the finite-size scaling (FSS) behavior for such a geometry. A general review of this subject is provided by reference \cite{Privman-89}. A detailed discussion thereof in the context of critical Casimir forces can be found in \cite{PTD-10}.

According to renormalization group (RG) theory \cite{Wegner-76}, close to the critical point and in the absence of an external  bulk field, the free-energy density $\cal F$ per $k_BT$ of the system (i.e., the free energy divided by $LL_\parallel^{2} k_BT$) decomposes into a singular contribution ${\cal F}^{\rm (s)}(t,L,L_\parallel)$ and a non-singular background term ${\cal F}^{\rm (ns)}(t,L,L_\parallel)$:
\begin{equation}
{\cal F}(t,L,L_\parallel) = {\cal F}^{\rm (s)}(t,L,L_\parallel) + {\cal F}^{\rm (ns)}(t,L,L_\parallel),
\label{free_sns}
\end{equation}
where $t\equiv (T-T_c)/T_c$ is the reduced temperature and $T_c$ is the bulk critical temperature. The non-singular background ${\cal F}^{\rm (ns)}$ decomposes further into specific geometric contributions, such as bulk, surface, and line terms which are regular functions of the Hamiltonian parameters and temperature; except for the bulk term, they depend on the {\bc}. Instead, the singular part of the free-energy density is a non-analytic function which exhibits a scaling behavior in the vicinity of the phase transition. In the FSS limit, i.e., in the limit $L$, $L_\parallel\rightarrow\infty$, $T\rightarrow T_c$ at fixed ratios $L/L_\parallel$ and $\xi/L$, where $\xi$ is the bulk correlation length, and neglecting corrections to scaling ${\cal F}^{\rm (s)}(t,L,L_\parallel)$ exhibits the following scaling property:
\begin{eqnarray}
\label{free_full_fss}
&{\cal F}^{\rm (s)}(t,L,L_\parallel) = \frac{1}{L^3}f\left(\tau,\rho,\ldots\right), \nonumber \\
&\tau \equiv \left(L/\xi_0^+\right)^{1/\nu}t, \qquad \rho \equiv L/L_\parallel,
\end{eqnarray}
where $f\left(\tau,\rho,\ldots\right)$ is a universal scaling function, $\nu$ is the  critical exponent of the bulk correlation length $\xi$, and $\xi_0^+$ is its nonuniversal amplitude:
\begin{equation}
\label{xi_crit}
\xi(t\rightarrow 0^\pm)=\xi_0^\pm|t|^{-\nu} =\xi_\pm.
\end{equation}
In \eeref{free_full_fss} the dots $\ldots$ denote the possible dependence of $f\left(\tau,\rho,\ldots\right)$ on certain additional scaling variables, the presence of which depends on the {\bc}; accordingly, the FSS limit is taken by keeping also such additional scaling variables fixed. The bulk free-energy density $f_{\rm bulk}(t)$ is obtained by taking the thermodynamic limit:
\begin{equation}
f_{\rm bulk}(t)\equiv \lim_{L,L_\parallel\rightarrow\infty}{\cal F}(t,L,L_\parallel)
\label{Fbulklimit}
\end{equation}
which is independent of the {\bc}. Analogously to \eeref{free_sns}, $f_{\rm bulk}(t)$ decomposes into a singular and a non-singular contribution:
\begin{equation}
f_{\rm bulk}(t)=f^{(s)}_{\rm bulk}(t)+f^{(ns)}_{\rm bulk}(t)
\end{equation}
with
\begin{eqnarray}
f^{(s)}_{\rm bulk}(t\to0^\pm) \nonumber \\
\label{fbulkcrit}
=k_BT\frac{LL_\parallel^{d-1}}{\xi_\pm^d}\frac{a_b^\pm}{\alpha(1-\alpha)(2-\alpha)}\left(k_BTLL_\parallel^{d-1}\right)^{-1}\\
=\left(\xi_0^\pm\right)^{-d}\frac{a_b^\pm}{\alpha(1-\alpha)(2-\alpha)}|t|^{d\nu}, \nonumber
\end{eqnarray}
where $\alpha=2-d\nu$ is the bulk critical exponent of the specific heat and $a_b^\pm$ are universal bulk amplitudes. The excess free energy $f_{\rm ex}^{\rm (s)}$ is defined as the remainder of the singular part of the free-energy density ${\cal F}^{\rm (s)}$ after subtraction of the bulk term $f^{\rm (s)}_{\rm bulk}(t)$:
\begin{equation}
\label{free_ex_def}
f_{\rm ex}^{\rm (s)}(t,L,L_\parallel)\equiv {\cal F}^{\rm (s)}(t,L,L_\parallel)-f^{\rm (s)}_{\rm bulk}(t).
\end{equation}
The critical Casimir force $F_C$ per area $L_\parallel^{(d-1)}$ and in units of $k_BT$ is defined as
\begin{equation}
\label{casimir_def}
F_C\equiv-\frac{\partial \left(Lf^{\rm (s)}_{\rm ex}\right)}{\partial L}\Bigg|_{t,L_\parallel}.
\end{equation}
By using \eeref{free_full_fss} in \eeref{casimir_def}, the critical Casimir force can be expressed as
\begin{equation}
\label{casimir_fss_leading}
F_C\left(t,L,L_\parallel\right)=\frac{1}{L^3}\theta\left( \tau= \left(L/\xi_0^+\right)^{1/\nu}t, \rho=L/L_\parallel, \ldots\right),
\end{equation}
where $\theta(\tau,\rho,\ldots)$ is a universal scaling function (compare \eeref{free_full_fss}). By using the asymptotic expression of \eeref{fbulkcrit} in \eeref{free_ex_def}, $\theta(\tau,\rho,\ldots)$ can be related to the scaling function $f(\tau, \rho, \ldots)$ of \eeref{free_full_fss} as
\begin{eqnarray}
\theta(\tau,\rho,\ldots) = &(d-1)f(\tau, \rho, \ldots) - \frac{\tau}{\nu}\frac{\partial f}{\partial\tau}(\tau, \rho, \ldots) \nonumber \\
&+\frac{a_b^\pm}{\alpha(1-\alpha)(2-\alpha)}\left(\frac{\xi_0^+}{\xi_0^\pm}\right)^d|\tau|^{d\nu},
\label{theta_xi0pm}
\end{eqnarray}
which holds for $\tau\rightarrow 0$. (The last expression in \eeref{theta_xi0pm} is not symmetric with respect to interchanging $+$ and $-$ because the scaling variable $\tau=\left(L/\xi_0^+\right)^{1/\nu}t$ is formed in terms of $L$ measured in units of $\xi_0^+$ both for $t>0$ and $t<0$.) The ratio $\xi_0^+/\xi_0^\pm$ appearing in \eeref{theta_xi0pm} is universal: it is equal to $1$ for $t>0$ and for $t<0$ it equals the universal ratio $\xi_0^+/\xi_0^-$ of the amplitudes of the correlation length. At the critical point $\tau=0$, the force is given by
\begin{equation}
F_C\left(t=0,L,L_\parallel\right)=\frac{1}{L^d}\Theta\left(\rho,\ldots\right),
\label{casimir_fss_leading_critical}
\end{equation}
with
\begin{equation}
\label{theta_Theta}
\Theta(\rho,\ldots)\equiv\theta(\tau=\ 0,\rho,\ldots).
\end{equation}
As in \eeref{free_full_fss}, in equations \eref{casimir_fss_leading} and \eref{theta_Theta} the additional dots $\ldots$ refer to possible additional scaling variables which enter into the FSS ansatz. Here and in the following we consider the film geometry only in the limit of small aspect ratios $\rho\rightarrow 0$; the full dependence of the critical Casimir force on the aspect ratio $\rho$ has been studied in \cite{HGS-11} for periodic {\bc}.

Finally, we observe that the scaling behavior reported in equations \eref{free_full_fss}, \eref{casimir_fss_leading}, and \eref{casimir_fss_leading_critical} is valid only up to corrections to scaling. The Monte Carlo results presented in \sref{sec:mc} have been obtained using an improved lattice model, in which the leading scaling corrections are suppressed \cite{PV-02}.

\subsection {Chemical-step contribution}
\label{sec:fss:cs}
\begin{figure}
\begin{center}
\includegraphics[clip=true,width=0.9\linewidth,keepaspectratio]{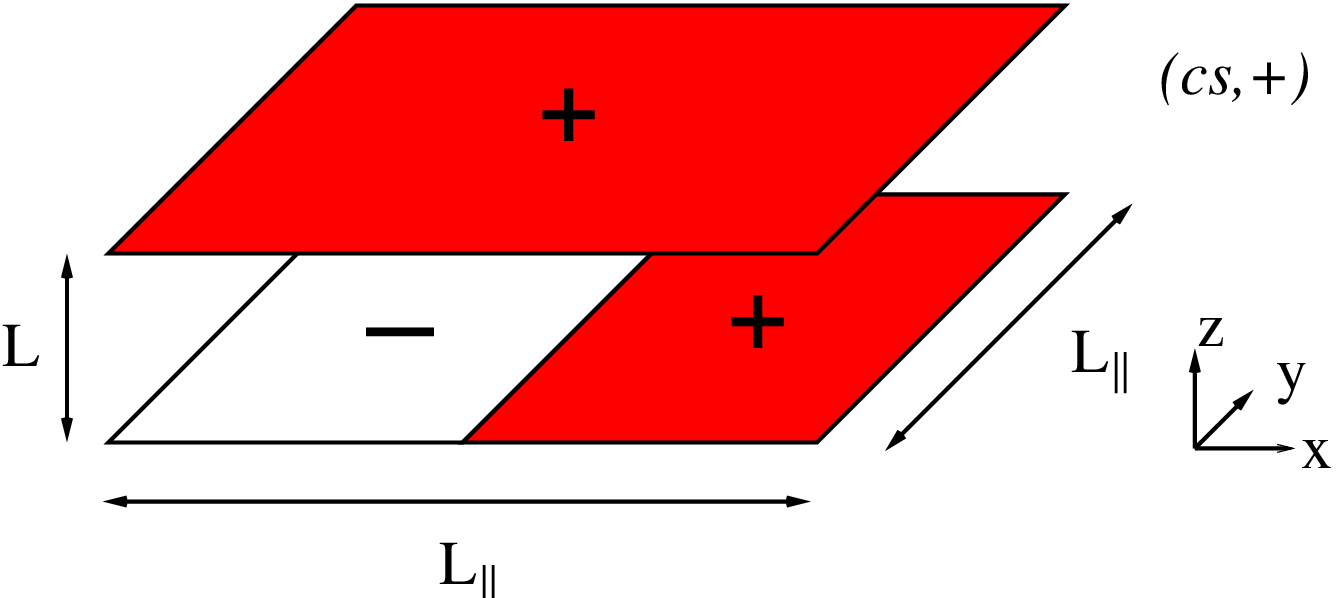}
\end{center}
\caption{Film geometry with aspect ratio $\rho=L/L_\parallel$ confined by a laterally homogeneous upper surface $(+)$ and by a lower surface with a chemical step $(cs)$.}
\label{cs}
\end{figure}
\begin{figure}
\begin{center}
\includegraphics[clip=true,width=0.9\linewidth,keepaspectratio]{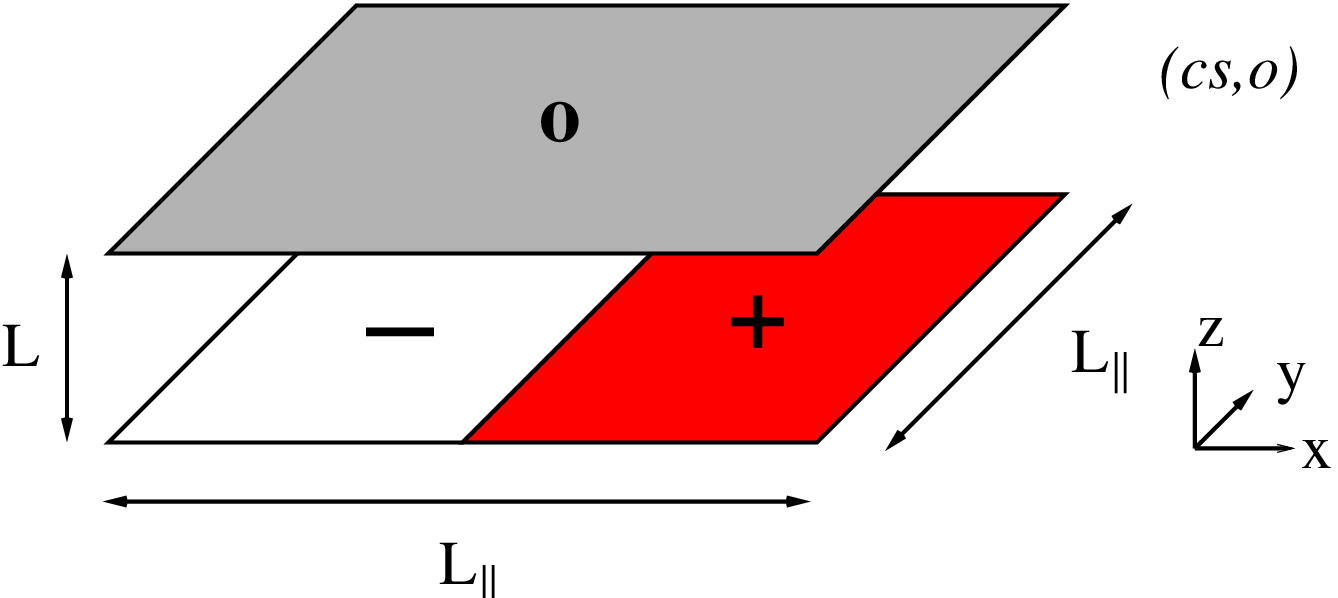}
\end{center}
\caption{Film geometry with aspect ratio $\rho=L/L_\parallel$ confined by a upper surface with open {\bc} $(o)$ and a by a lower surface with a chemical step $(cs)$.}
\label{csopen}
\end{figure}
In order to determine line contributions to the critical Casimir force, we consider the {\bc} illustrated in figures \ref{cs} and \ref{csopen} as the simplest realization of the film geometry in the presence of a chemical step. We divide the lower surface into two halves, each of them characterized by a homogeneous but opposite adsorption preference, such that the system remains translationally invariant along the y-direction (figures \ref{cs} and \ref{csopen}) and displays a straight chemical step $(cs)$ separating the two halves. We note that the lateral periodic {\bc} in the x-direction induces an additional, second chemical step at the lateral boundaries. For the opposing surface we choose a homogeneous adsorption preference $(+)$ (\fref{cs}) or open $(o)$ {\bc} (\fref{csopen}). We refer to these {\bc} as $(cs,+)$ and $(cs,o)$, respectively. In the slab limit $\rho\rightarrow 0$ the critical Casimir force for $(cs,+)$ {\bc} reduces to the mean value of the force for a system in which both confining surfaces exhibit the same homogeneous adsorption preference corresponding to the so-called $(+,+)$ {\bc}, and of the force for a system in which the walls have the opposite homogeneous adsorption preference corresponding to the so-called  $(+,-)$ {\bc}. Analogously, in the limit $\rho\rightarrow 0$, the critical Casimir force for $(cs,o)$ {\bc} reduces to the mean value of the force for a system in which one of the confining surfaces exhibits a homogeneous adsorption preference while the opposite wall has open {\bc} corresponding to the so-called $(+,o)$ {\bc}, and of the force for the same system, in which the adsorption preference is opposite corresponding to the so-called $(-,o)$ {\bc}. In the absence of a bulk field, these two {\bc} are effectively identical, so that we conclude that for $\rho\rightarrow 0$ the critical Casimir force for $(cs,o)$ {\bc} approaches the one for $(+,o)$ {\bc}.

The above scaling properties are essentially due to the fact that, for a confined system with the lateral critical fluctuations not fully developed (i.e., $T\ne T_c(L)$), the correlation length in the film is bounded by the slab thickness $L$. Therefore the chemical steps represent line defects the contribution of which to the critical Casimir force per area vanishes in the limit of large lateral size $L_\parallel\rightarrow \infty$ (see the discussion at the end of section 3 in \cite{PTD-10}). Accordingly, one expects that the presence of two chemical steps enters into the dependence of the critical Casimir force on the aspect ratio $\rho$.

In the following we provide a summary of the arguments which allow one to formalize this concept. A detailed discussion thereof is provided in \cite{PTD-10}.
For $(cs,+)$ {\bc} we define the chemical-step contribution to the critical Casimir force $F_{C,steps}$ as
\begin{eqnarray}
F_{C,steps}&(t,L,L_\parallel) \equiv F_C(t,L,L_\parallel) \nonumber\\
&- \frac{F_{C,(+,+)}(t,L,L_\parallel) + F_{C,(+,-)}(t,L,L_\parallel)}{2},
\label{Fcsplus}
\end{eqnarray}
where $F_{C,(+,+)}(t,L,L_\parallel)$ and $F_{C,(+,-)}(t,L,L_\parallel)$ are the critical Casimir forces for laterally homogeneous $(+,+)$ and $(+,-)$ {\bc}, respectively. For $(cs,o)$ {\bc} we define $F_{C,steps}$ as
\begin{equation}
F_{C,steps}(t,L,L_\parallel) \equiv F_C(t,L,L_\parallel) - F_{C,(+,o)}(t,L,L_\parallel),
\label{Fcsopen}
\end{equation}
where $F_{C,(+,o)}(t,L,L_\parallel)$ is the critical Casimir force for laterally homogeneous $(+,o)$ {\bc}. The definitions of $F_{C,steps}$ given in equations \eref{Fcsplus} and \eref{Fcsopen} correspond to the force per area $L_\parallel^{d-1}$ which remains if one subtracts the mean value of the forces per area (keeping $L_\parallel<\infty$) for laterally homogeneous {\bc} obtained by considering separately the two lower halves which form the chemical steps (see figures \ref{cs} and \ref{csopen}); this mean value is the force per area which is expected if the chemical steps would not give rise to a contribution to $F_C$.

{\it Off criticality}, and in the limit $L$, $L_\parallel\rightarrow \infty$ {\em at fixed $T$}, the reduced free-energy density for $(cs,+)$ and $(cs,o)$ {\bc} decomposes as
\begin{eqnarray}
{\cal F}(t,L,L_\parallel) = f_{\rm bulk}&(t) + \frac{1}{L} f_{\rm surf}(t) \nonumber \\
&+ \frac{\rho}{L^2} f_ {\rm steps}(t) + \Or(e^{-L/\xi}/L).
\label{offdecomposition}
\end{eqnarray}
For the free energy ${\cal F}^*=k_BTLL_\parallel^2{\cal F}$ \eeref{offdecomposition} implies ${\cal F}^*=k_BT\{LL_\parallel^2f_{\rm bulk}(t)+L_\parallel^2f_{\rm surf}(t)+L_\parallel f_{\rm steps}(t) + \Or(e^{-L/\xi}L_\parallel^2)\}$ so that $k_BTL_\parallel^2f_{\rm surf}(t)$ and $k_BTL_\parallel f_{\rm steps}(t)$ can be identified as the surface and the line contribution, respectively, to the free energy where $f_{\rm steps}(t)$ is generated by the two individual chemical steps. However, in the FSS limit the decomposition given by \eeref{offdecomposition} becomes blurred. As mentioned above, the non-singular part of the free energy is expected to display a geometrical decomposition analogous to \eeref{offdecomposition}. Instead, concerning the singular part of the free energy a priori one cannot identify a surface or a line term. Nevertheless, generalizing the discussion in \cite{Privman-89}, one can formally define a line free energy by comparing the free energy for $(cs,+)$ and $(cs,o)$ {\bc} with the ones for suitable reference systems, which do not exhibit a line defect. Accordingly, we define $\hat{f}_{\rm steps}(t,L)$ as
\begin{eqnarray}
\hat{f}_ {\rm steps}(t,L)\equiv L^2\Bigg(\frac{\partial}{\partial\rho}\Big|_{L,t}\Big[&{\cal F}(t,L,L_\parallel) \nonumber \\
&-{\cal F}^{\rm (ref)}(t,L,L_\parallel)\Big]\Bigg)\Bigg|_{\rho=0}.
\label{edge}
\end{eqnarray}
In agreement with \eeref{Fcsplus}, as reference free energy for $(cs,+)$ {\bc} we take
\begin{eqnarray}
{\cal F}^{\rm (ref)}(t,L,L_\parallel) = \frac{1}{2}\Big[&{\cal F}_{(+,+)}(t,L,L_\parallel) \nonumber \\
&+{\cal F}_{(+,-)}(t,L,L_\parallel)\Big],
\label{refcsplus}
\end{eqnarray}
where ${\cal F}_{(+,+)}$ and ${\cal F}_{(+,-)}$ are the free-energy densities per $k_BT$ for $(+,+)$ and $(+,-)$ {\bc}, respectively. In line with \eeref{Fcsopen}, for $(cs,o)$ {\bc} we take
\begin{equation}
{\cal F}^{\rm (ref)}(t,L,L_\parallel) = {\cal F}_{(+,o)}(t,L,L_\parallel),
\label{refcsopen}
\end{equation}
i.e., the free-energy density per $k_BT$ for $(+,o)$ {\bc}. The choice of the definitions given by equations \eref{refcsplus} and \eref{refcsopen} ensures that ${\cal F}(t,L,L_\parallel)$ coincides with ${\cal F}^{\rm (ref)}(t,L,L_\parallel)$ in the limit $\rho\rightarrow 0$. Using these definitions it follows that (see equations \eref{Fcsplus}, \eref{Fcsopen}, and \eref{edge})
\begin{eqnarray}
F_{C,{\rm steps}}(t,L,L_\parallel)= &-\frac{1}{L_\parallel}\left(\frac{\partial \hat{f}^{\rm (s)}_{\rm steps}(t,L)}{\partial L}\Bigg|_t\right) \nonumber \\
&+ o(\rho), \qquad \rho\rightarrow 0.
\label{forcecs}
\end{eqnarray}
\Eref{forcecs} shows that, indeed, the chemical-step contribution to the critical Casimir force is solely due to the line free energy $\hat{f}_ {\rm steps}(t,L)$ (see \eeref{edge}). Moreover, using the definition of $\hat{f}_ {\rm steps}(t,L)$, one can show that the first term on the rhs of \eeref{forcecs} is $\propto\rho$. In the case of homogeneous {\bc} and in the absence of a phase transition at $T_c(L)<T_c(L=\infty)$ associated with a divergent lateral correlation length, for $\rho\rightarrow 0$ the dependence of the free energy and of the critical Casimir force on the aspect ratio $\rho$ is rather weak. In fact, it is expected to be exponentially small on the scale of the lateral correlation length. This holds in the case of $(+,+)$, $(+,-)$, and $(+,o)$ {\bc} which in equations \eref{refcsplus} and \eref{refcsopen} form the reference systems for $(cs,+)$ and $(cs,o)$ {\bc}. (We mention that even for $(o,o)$ {\bc}, for which the presence of the effectively two-dimensional ordering transition at $T=T_c(L)$ results in the onset of a diverging lateral correlation length, the critical Casimir force is found to display only a weak aspect ratio dependence, which furthermore is limited to a narrow interval in $\tau$ \cite{PTTD-13}.) Thus, we are lead to conclude that the aspect ratio dependence of the critical Casimir force for the reference systems with $(+,+)$, $(+,-)$, and $(+,o)$ {\bc} is negligible for $\rho\rightarrow 0$. Therefore it is justified to assume that for these {\bc} the dependence of the critical Casimir force on $\rho$ is at least quadratic in $\rho$ for $\rho\rightarrow 0$, that is,
\begin{eqnarray}
&\theta_{(+,+)}(\tau,\rho)-\theta_{(+,+)}(\tau,\rho=0)=o(\rho), \quad &\rho\rightarrow 0,\nonumber \\
\label{osmall}
&\theta_{(+,-)}(\tau,\rho)-\theta_{(+,-)}(\tau,\rho=0)=o(\rho), \quad &\rho\rightarrow 0, \\
&\theta_{(+,o)}(\tau,\rho)-\theta_{(+,o)}(\tau,\rho=0)=o(\rho), \quad &\rho\rightarrow 0.\nonumber 
\end{eqnarray}
\Eref{osmall} allows one to express \eeref{forcecs} as
\begin{equation}
F_{C,{\rm steps}}(t,L,L_\parallel) = \frac{\rho}{L^3}\frac{\partial\theta(\tau,\rho,\ldots)}{\partial\rho}\Bigg|_{\rho=0} + o(\rho).
\label{forcecsrhoderivative}
\end{equation}
\Eref{forcecsrhoderivative} suggests to introduce a Taylor expansion of the universal scaling function $\theta(\tau,\rho,\ldots)$ (see \eeref{casimir_fss_leading}) for its dependence on $\rho\rightarrow 0$:
\begin{equation}
\theta(\tau,\rho,\ldots) = \theta(\tau,\rho=0,\ldots) + \rho E(\tau,\ldots) + o(\rho),
\label{thetaexp}
\end{equation}
so that \eeref{forcecsrhoderivative} can be rewritten as
\begin{equation}
F_{C,{\rm steps}}(t,L,L_\parallel) = \frac{\rho}{L^3}E(\tau)+ o(\rho).
\label{forcecsrho}
\end{equation}
Taking into account \eeref{osmall}, \eeref{forcecsrho} implies that for $\rho\ll 1$ that contribution to $F_C$, which is linear in $\rho$, is solely due to the presence of the chemical steps on the lower surfaces and thus serves as their fingerprint on the critical Casimir force.

We observe that in equations \eref{forcecsrhoderivative} and \eref{forcecsrho} the identification of the chemical-step contribution, as that part of $\theta(\tau,\rho,\ldots)$ which is linear in $\rho$, is always done up to possible higher-order terms in $\rho$. This is an unavoidable consequence of the definitions used here, and in particular of the fact that we define $F_{C,{\rm steps}}$ in {equations \eref{Fcsplus} and \eref{Fcsopen}} for an arbitrary aspect ratio $\rho$. On the other hand, the interpretation of the critical Casimir force for these {\bc} in terms of individual chemical steps is reasonable for $\rho\ll 1$ only: for a sufficiently large aspect ratio $\rho$ the decomposition of the critical Casimir force as the sum of the force in the slab limit $\rho\rightarrow 0$ and of the contribution of two chemical steps breaks down. This is analogous to the geometrical decomposition of the singular part of the free-energy density which becomes blurred in the scaling region. In this sense, we can identify $E(\tau,\ldots)$ as the chemical-step contribution to $\theta(\tau,\rho,\ldots)$ in the region $0<\rho\ll 1$ where any higher-order terms in $\rho$ are negligible. For the sake of brevity, in the following we neglect the possible corrections $o(\rho)$ to the critical Casimir force.

\begin{figure}
\begin{center}
\includegraphics[clip=true,width=0.9\linewidth,keepaspectratio]{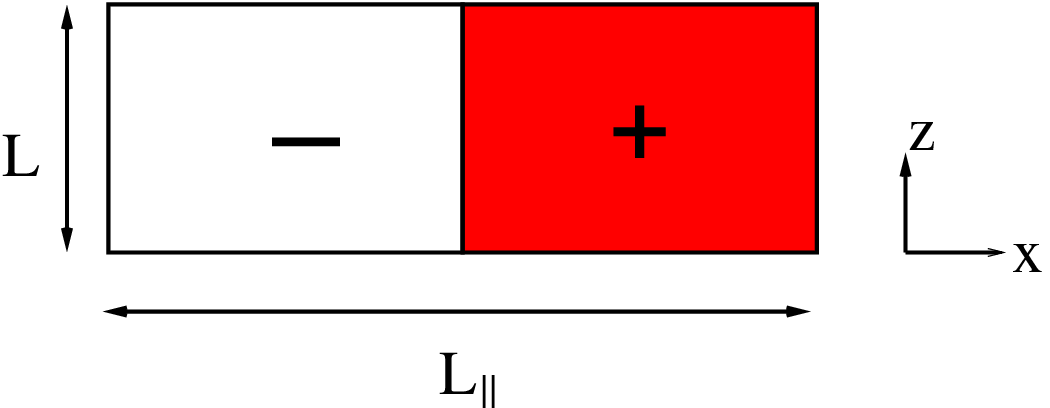}
\end{center}
\caption{A section of the ground-state configuration at $y=const$ for the {\bc} shown in \fref{csopen}. The configuration is translationally invariant along the $y$ direction. Due to the periodic {\bc} in the lateral directions, the configuration consists of two interfaces of area $L\times L_\parallel$ each, both perpendicular to the substrate.}
\label{csopen_gs}
\end{figure}
For $(cs,o)$ {\bc} and in the limit $\tau\rightarrow -\infty$, the system approaches the ground-state configuration. Simple energy considerations allow one to determine the ground state configuration as the one consisting of two interfaces, aligned with the underlining chemical surface pattern and perpendicular to the confining surfaces. This configuration is illustrated in \fref{csopen_gs}. In this case the chemical-step contribution to the critical Casimir force is determined by the interfacial tension. It is given by $-R_\sigma|\tau|^\mu$, where $R_\sigma= \sigma_0(\xi_0^+)^{d-1}/(k_BT_c)$ is the universal amplitude ratio for the interfacial tension $\sigma=\sigma_0|t|^\mu$ associated with the spatially coexisting bulk phases; {$\mu=(d-1)\nu=1.26004(20)$ \cite{Hasenbusch-10} for $d=3$} is its critical exponent. The universal amplitude ratio $R_\sigma$ has been determined for the three-dimensional Ising UC as $R_\sigma=0.377(11)$ \cite{ZF-96} and, more recently, as $R_\sigma=0.387(2)$ \cite{CHP-07}. Thus for $(cs,o)$ {\bc} and in the limit $\tau\rightarrow -\infty$ the scaling function $E_{(cs,o)}(\tau)$ approaches
\begin{equation}
E_{(cs,o)}(\tau\ll-1)\simeq -2R_\sigma|\tau|^{\mu},
\label{Ecsopen_interface}
\end{equation}
where the factor $2$ accounts for the presence of actually two interfaces.

\subsection{Chemical-step contributions as building blocks}
\label{sec:fss:blocks}
\begin{figure}
\begin{center}
\includegraphics[clip=true,width=0.9\linewidth,keepaspectratio]{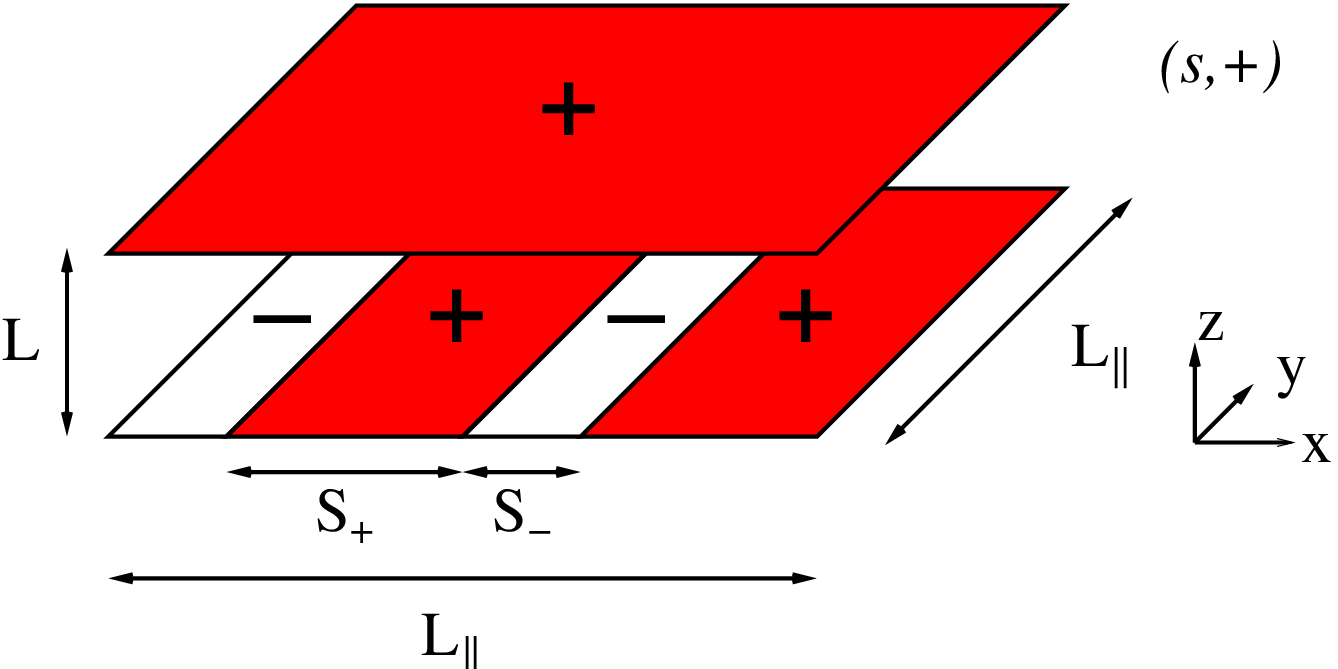}
\end{center}
\caption{Film geometry confined by a laterally homogeneous upper surface and by a chemically striped bottom substrate with stripes of alternating adsorption preference; $\kappa=S_+/L$ and $\varsigma=S_+/S_-$.}
\label{bcstripes}
\end{figure}
\begin{figure}
\begin{center}
\includegraphics[clip=true,width=0.9\linewidth,keepaspectratio]{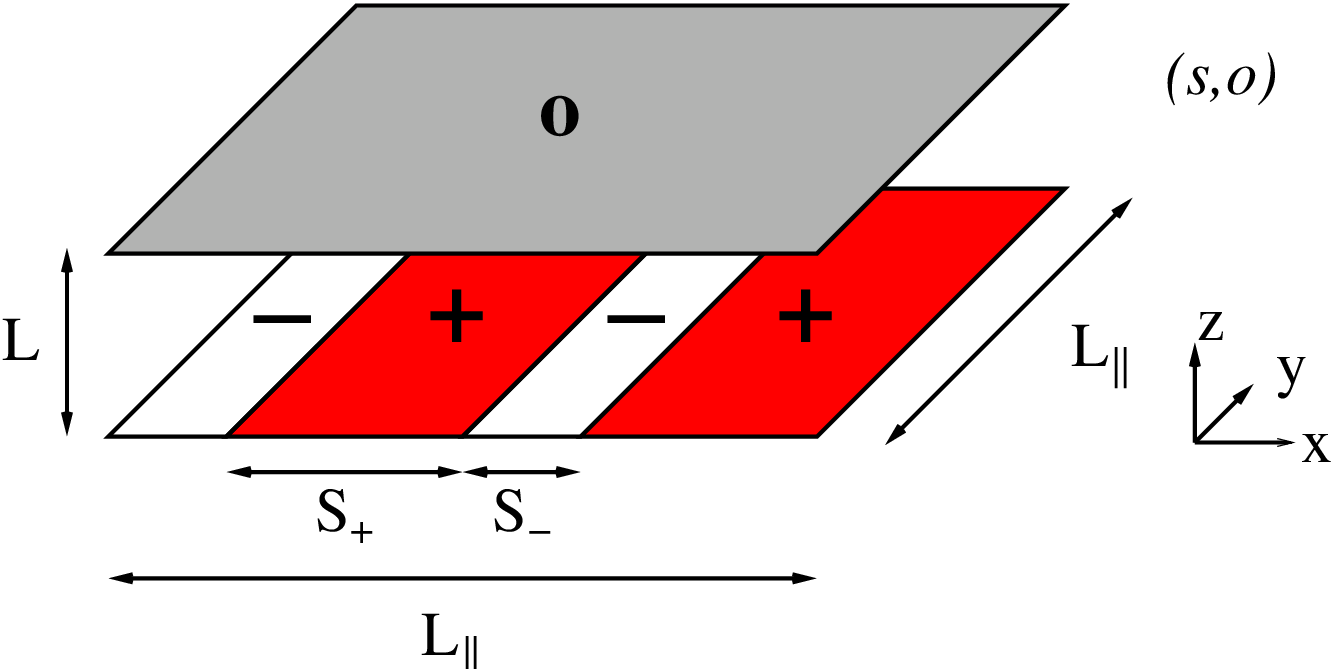}
\end{center}
\caption{Film geometry confined by a laterally homogeneous upper surface with open {\bc} and by a chemically striped bottom substrate with stripes of alternating adsorption preference; $\kappa=S_+/L$ and $\varsigma=S_+/S_-$.}
\label{bcstripesopen}
\end{figure}
The knowledge of the universal scaling function $E(\tau)$ is not only of interests {\it per se}, as it elucidates the FSS behavior of confined systems in the presence of a chemical step, but in the limit of wide stripes it also serves as a building block for computing the critical Casimir force for more complex chemically-striped {\bc}. To be specific, we consider a chemically striped substrate, consisting of straight stripes of widths $S_+$ and $S_-$ with alternating adsorption preferences. For the opposing surface, we choose a homogeneous adsorption preference or open {\bc}. In figures \ref{bcstripes} and \ref{bcstripesopen} we illustrate these {\bc}; we refer to them as $(s,+)$ and $(s,o)$, respectively. The presence of two additional lengths $S_+$ and $S_-$ results in the dependence of the critical Casimir force on two additional scaling variables: $\kappa\equiv S_+/L$ and $\varsigma\equiv S_+/S_-$. Furthermore, we restrict the discussion of the critical Casimir force for $(s,+)$ and $(s,o)$ {\bc} to the slab limit $\rho=0$. In the limiting case $\varsigma\rightarrow 0$, the {\bc} $(s,+)$ and $(s,o)$ reduce to $(+,-)$ and $(+,o)$ {\bc}, respectively, whereas in the limiting case $\varsigma\rightarrow\infty$, the {\bc} $(s,+)$ and $(s,o)$ reduce to $(+,+)$ and $(+,o)$ {\bc}, respectively. For fixed $0<\varsigma<\infty$ and in the limit $\kappa\rightarrow \infty$, the stripes become well separated, so that the critical Casimir force reduces to the sum of the force for single stripes and of the contribution of the chemical steps separating the stripes. For the present geometry one has $N_{\rm steps}=2L_\parallel/(S_++S_-)$ such steps, each of them giving rise to a contribution to the critical Casimir force proportional to the aspect ratio $\rho=L/L_\parallel$. (Note that in the limit $L_\parallel\rightarrow\infty$ with $L$, $S_+$, and $S_-$ fixed the aspect ratio $\rho=L/L_\parallel$ vanishes and the number of steps $N_{\rm steps}$ diverges such that their product $\rho N_{\rm steps}$ attains the finite and nonzero value $2\varsigma/(\kappa(1+\varsigma))$.) The asymptotic behavior for $\kappa \gg 1$ of the universal scaling function $\theta_{(s,+)}(\tau,\rho=0,\kappa,\varsigma)$ for $(s,+)$ {\bc} is therefore given by
\begin{eqnarray}
\theta_{(s,+)}(\tau, \rho\rightarrow 0,\kappa\gg 1,0<\varsigma<\infty) \nonumber \\
=\frac{S_+\theta_{(+,+)}(\tau)+S_-\theta_{(+,-)}(\tau)}{S_++S_-} + \frac{1}{2}\rho N_{\rm steps} E_{(cs,+)}(\tau) \nonumber \\
=\frac{\varsigma\theta_{(+,+)}(\tau)+\theta_{(+,-)}(\tau)}{\varsigma+1} + \frac{\varsigma}{\kappa(1+\varsigma)}E_{(cs,+)}(\tau).
\label{thetasp}
\end{eqnarray}
(Note that the scaling function $E(\tau)$ as introduced in equations \eref{thetaexp} and \eref{forcecsrho} holds for the geometries shown in figures \ref{bcstripes} and \ref{bcstripesopen} which due to the periodic lateral {\bc} contain {\it de facto} two chemical steps; this leads to the factor $\case{1}{2}$ in the first part of \eeref{thetasp}.) Analogously, the asymptotic behavior of the universal scaling function $\theta_{(s,o)}(\tau,\rho=0,\kappa,\varsigma)$ for $(s,o)$ {\bc} is
\begin{eqnarray}
\theta_{(s,o)}&(\tau,\rho=0,\kappa\gg 1,0<\varsigma<\infty) \nonumber \\
&=\theta_{(+,o)}(\tau) + \frac{\varsigma}{\kappa(1+\varsigma)}E_{(cs,o)}(\tau).
\label{thetaso}
\end{eqnarray}
In \cite{PTTD-13} we have found that \eeref{thetasp} is quantitatively reliable for $\varsigma=1$, $\kappa \gtrsim 2$, and a wide range of $\tau$ around the critical point $\tau=0$. (In order to check the reliability of this \eeref{thetasp}, in \cite{PTTD-13} we had used the scaling function $E_{(cs,+)}(\tau)$ as determined in \cite{PTD-10}.)

The prediction for the asymptotic behavior of the universal scaling functions given in equations \eref{thetasp} and \eref{thetaso} can be easily generalized to an arbitrarily chemically striped substrate, provided that the widths of the stripes are large relative to the film thickness $L$.

\section{Monte Carlo results}
\label{sec:mc}
\subsection{Model and method}
\label{sec:mc:model}
In order to compute the critical Casimir force for a confined binary liquid mixture close to its bulk critical  demixing point, we have performed MC simulations for a lattice Hamiltonian representing the 3D Ising universality class. In accordance with previous numerical studies \cite{Hasenbusch-10c,PTD-10,Hasenbusch-11,Hasenbusch-12,Hasenbusch-12b,PTTD-13,PT-13}, we have studied the so-called improved Blume-Capel model \cite{Blume-66,Capel-66}. It is defined on a three-dimensional simple cubic lattice, with a spin variable $S_i$ on each site $i$ which can take the values $S_i=-1$, $0$, $1$. The Hamiltonian of the model per $k_BT$ is
\begin{equation}
\label{bc}
{\cal H}=-\beta\sum_{<i j>}S_i S_j + D\sum_i S_i^2,\qquad S_i=-1,0,1,
\end{equation}
so that the Gibbs weight is $\exp(-\cal H)$ leading to the partition function
\begin{equation}
\label{Z}
Z(\beta,L,L_\parallel)\equiv\sum_{\{S_i\}} \exp(-{\cal H}),
\end{equation}
where $\{S_i\}$ is the configuration space of the Hamiltonian  given in \eeref{bc}. The partition function in \eeref{Z} depends implicitly also on the {\bc}. In line with the convention used in \cite{Hasenbusch-10,Hasenbusch-10c,PTD-10,PTTD-13,PT-13}, in the following we shall keep $D$ constant, considering it as a part of the integration measure over $\{S_i\}$, while we vary the coupling parameter $\beta$, which is proportional to the inverse temperature, $\beta\sim 1/T$. In the limit $D\rightarrow -\infty$ the configurations involving vacancies $S=0$ are suppressed and the Hamiltonian reduces to the one of the Ising model. For $d\ge 2$, starting from $D=-\infty$ the phase diagram of the model displays a line of second order phase transitions at $\beta_c=\beta_c(D)$ which reaches a tricritical point at $D=D_{\rm tri}$, beyond which the phase transition is of first order. The value of $D_{\rm tri}$ in $d=3$ has been determined as $D_{\rm tri}=2.006(8)$ in \cite{Deserno-97}, as $D_{\rm tri}\simeq 2.05$ in \cite{HB-98}, and more recently as $D_{\rm tri}=2.0313(4)$ in \cite{DB-04}. At $D=0.656(20)$ \cite{Hasenbusch-10} the model is improved, i.e., the leading corrections to scaling $\propto L^{-\omega}$ with $\omega=0.832(6)$ \cite{Hasenbusch-10} are suppressed. In the MC results presented here, $D$ is fixed at $D=0.655$, which is the value of $D$ used in most of the recent simulations of the improved Blume-Capel model \cite{Hasenbusch-10c,Hasenbusch-11,Hasenbusch-12b,Hasenbusch-10,PTTD-13,PT-13}. For this value of the reduced coupling $D$ the model is critical for $\beta=\beta_c=0.387721735(25)$ \cite{Hasenbusch-10}.

Specificly, we consider a three-dimensional simple cubic lattice $L\times L_\parallel\times L_\parallel$ with periodic {\bc} in the lateral directions $x$ and $y$. The lattice constant is set to $1$, i.e., here dimensionless lengths have to be multiplied with the lattice constant in order to become physical lengths. For the two confining surfaces we impose the {\bc} shown in figures \ref{cs} and \ref{csopen}. The {\bc} illustrated in \fref{cs} are realized by fixing the spins at the two surfaces $z=0$ and $z=L$, so that there are $L-1$ layers of fluctuating spins. The spins at the upper surface $z=L$ are fixed to $+1$, as to realize a homogeneous adsorption preference, while the lower surface $z=0$ is divided into two halves, one with spins fixed to $-1$ and the other half with spins fixed to $+1$. In the case of $(cs,o)$ {\bc} the spins on the lower surface $z=0$ are fixed in the same way as for $(cs,+)$ {\bc}, while we employ open {\bc} on the upper surface, so that in this case there are $L$ layers of fluctuating spins. This definition of $L$ assigns a common thickness to films for both $(cs,+)$ and $(cs,o)$ {\bc}. It also facilitates the comparison with the mean field calculations presented in \sref{sec:mft} because the lattice thickness $L$ of the films here is the actual continuum thickness divided by the lattice constant.

The numerical determination of the critical Casimir force proceeds by replacing the derivative in \eeref{casimir_def} by a finite difference $\Delta{\cal F}$ between the free energies ${\cal F}$ (divided by $k_BTLL_\parallel^2$) of a film of thickness $L$ and of a film of thickness $L-1$:
\begin{equation}
\Delta{\cal F}(t,L,L_\parallel) \equiv L{\cal F}(t,L,L_\parallel) - (L-1){\cal F}(t,L-1,L_\parallel).
\label{deltaF}
\end{equation}
Using equations \eref{Fbulklimit} and \eref{free_ex_def}, the definition of the critical Casimir force given in \eeref{casimir_def}, and its leading scaling behavior provided in \eeref{casimir_fss_leading}, $\Delta{\cal F}(t,L,L_\parallel)$ can be expressed as
\begin{eqnarray}
\Delta{\cal F}&(t,L,L_\parallel) = f_{\rm bulk}(t)-\frac{1}{(L-1/2)^3}\cdot \nonumber \\
&\theta\Bigg(\tau=\Bigg(\frac{L-1/2}{\xi_0^+}\Bigg)^{1/\nu}t, \rho=\frac{L-1/2}{L_\parallel}\Bigg),
\label{deltaF_leading}
\end{eqnarray}
where for $(cs,+)$ and $(cs,o)$ {\bc} studied here there are no additional scaling variables. The meaning of \eeref{deltaF_leading} is that on the rhs one has $\theta(\tau,\rho)$ evaluated at $\tau =\left((L-1/2)/\xi_0^+\right)^{1/\nu}t$ and $\rho=(L-1/2)/L_\parallel$, respectively. Note that in \eeref{deltaF} the non-singular parts of the surface and of the line free energy drop out from the free energy difference. In fact, the non-singular part of the free energy exhibits a geometrical decomposition analogous to \eeref{offdecomposition} \cite{Privman-89}. By using such a decomposition, the non-singular part of the free energy difference of \eeref{deltaF} can be expressed as
\begin{eqnarray}
&L{\cal F}^{\rm (ns)}(t,L,L_\parallel) - (L-1){\cal F}^{\rm (ns)}(t,L-1,L_\parallel) \nonumber \\
&=L\frac{LL_\parallel^2f^{\rm (ns)}_{\rm bulk}(t) + L_\parallel^2f^{\rm (ns)}_{\rm surf}(t) + L_\parallel f^{\rm (ns)}_{\rm steps}(t)}{LL_\parallel^2} \nonumber \\
&-(L-1)\frac{(L-1)L_\parallel^2f^{\rm (ns)}_{\rm bulk}(t) + L_\parallel^2f^{\rm (ns)}_{\rm surf}(t) + L_\parallel f^{\rm (ns)}_{\rm steps}(t)}{(L-1)L_\parallel^2} \nonumber \\
&=f^{\rm (ns)}_{\rm bulk}(t).
\end{eqnarray}
Thus knowledge of $f_{\rm bulk}(t)$ and MC data for $\Delta{\cal F}$ render the universal scaling function $\theta$ of the critical Casimir force (\eeref{casimir_fss_leading}). These results for the critical Casimir force correspond to the intermediate film thickness $L-1/2$. This choice ensures that in the FSS limit no additional scaling corrections $\propto L^{-1}$ are generated \cite{PTD-10}. By inserting the expansion according to \eeref{thetaexp} into \eeref{deltaF_leading}, $\Delta{\cal F}$ can be related to the scaling function $E(\tau)$ which characterizes the chemical-step contribution to the critical Casimir force:
\begin{eqnarray}
\Delta{\cal F}(t,L,L_\parallel) = f_{\rm bulk}(t)\nonumber \\
-\frac{1}{(L-1/2)^3}\Bigg[\theta\Bigg(\tau=\Bigg(\frac{L-1/2}{\xi_0^+}\Bigg)^{1/\nu}t,\rho=0\Bigg) \nonumber \\
\qquad\qquad \ +\frac{L-1/2}{L_\parallel} E\Bigg(\tau=\Bigg(\frac{L-1/2}{\xi_0^+}\Bigg)^{1/\nu}t\Bigg) \Bigg].
\label{deltaF_leading_exp}
\end{eqnarray}
The scaling behaviors given by equations \eref{deltaF_leading} and \eref{deltaF_leading_exp} are valid up to corrections to scaling. In the Ising universality class, the leading scaling correction is due to the leading irrelevant bulk operator and is $\propto L^{-\omega}$ with $\omega=0.832(6)$ \cite{Hasenbusch-10}. As mentioned above, in the improved model considered here the amplitude of this correction to scaling is suppressed. In this case, the leading scaling correction stems from the {\bc}. Any {\bc} which are not fully periodic (or antiperiodic) give rise to scaling corrections $\propto L^{-1}$. As proposed first in \cite{CF-76}, such scaling corrections can be absorbed by the substitution $L\rightarrow L+c$, where $c$ is a nonuniversal temperature-independent length. This result can be interpreted within the framework of the so-called non-linear scaling fields \cite{AF-83}: while for periodic {\bc} $L$ is a scaling field by itself, for non-periodic {\bc} or, more generally, in the absence of translational invariance, $L$ has to be replaced by an analytic expansion, the leading term of which is $L+c$. This property has been confirmed in many numerical simulations of classical models \cite{Hasenbusch-08,Hasenbusch-09,Hasenbusch-09b,Hasenbusch-10c,PTD-10,Hasenbusch-11,Hasenbusch-12,Hasenbusch-12b,PTTD-13,PT-13} and it has been pointed out to hold also for FSS at a quantum phase transition \cite{CPV-14}. With the substitution $L\rightarrow L+c$ in equations \eref{deltaF_leading} and \eref{deltaF_leading_exp} one has
\begin{eqnarray}
\Delta{\cal F}(t,L,L_\parallel) = f_{\rm bulk}(t)-\frac{1}{(L-1/2+c)^3}\cdot \nonumber \\
\ \theta\left(\tau =\left(\frac{L-1/2+c}{\xi_0^+}\right)^{1/\nu}t,\rho=\ \frac{L-1/2+c}{L_\parallel}\right),
\label{deltaF_full}
\end{eqnarray}
and
\begin{eqnarray}
\Delta{\cal F}(t,L,L_\parallel) = f_{\rm bulk}(t)\nonumber \\
-\frac{1}{(L-1/2+c)^3}\Bigg[\theta\left(\tau =\left(\frac{L-1/2+c}{\xi_0^+}\right)^{1/\nu}t,\rho=0\right) \nonumber \\
+ \frac{L-1/2+c}{L_\parallel} E\left(\tau =\left(\frac{L-1/2+c}{\xi_0^+}\right)^{1/\nu}t\right) \Bigg],
\label{deltaF_full_exp}
\end{eqnarray}
respectively. Equations \eref{deltaF_full} and \eref{deltaF_full_exp} correspond to the FSS ansatz which we use for analyzing the MC data. A more detailed discussion on the corrections-to-scaling and possible modifications of equations \eref{deltaF_full} and \eref{deltaF_full_exp} can be found in \cite{PTTD-13}.

\subsection{Critical Casimir amplitude at $T_c$}
\label{sec:mc:tc}
In order to determine the critical Casimir force at $T_c$, we have computed the free energy difference $\Delta{\cal F}$ in equations \eref{deltaF_full} and \eref{deltaF_full_exp} by using the coupling parameter approach introduced in \cite{VGMD-07} and also used in Refs.~\cite{VGMD-08,PTD-10,VMD-11,PTTD-13,VED-13,VD-13,Vasilyev-14}, which we briefly describe here. Given two reduced Hamiltonians ${\cal H}_1$ and ${\cal H}_2$ sharing the same configuration space $\{C\}$ we consider the convex combination ${\cal H}(\lambda)$
\begin{equation}
{\cal H}(\lambda) \equiv \left(1-\lambda\right){\cal H}_1 + \lambda{\cal H}_2,\qquad \lambda\in \left[0,1\right].
\label{crossover_H}
\end{equation}
This Hamiltonian ${\cal H}(\lambda)$ (in units of $k_BT$) leads to a free energy ${\mathrm F}(\lambda)$ in units of $k_BT$\footnote{Note that the free energy ${\mathrm F}(\lambda)$ in units of $k_BT$ differs from the free energy density ${\cal F}(t,L,L_\parallel)$ introduced in \eeref{free_sns}, which is the free energy per volume in units of $k_BT$.}. Its derivative is
\begin{equation}
\frac{\partial {\mathrm F}(\lambda)}{\partial\lambda}=\frac{\sum_{\{C\}}\frac{\partial {\cal H}(\lambda)}{\partial\lambda}e^{-{\cal H}(\lambda)}}{\sum_{\{C\}}e^{-{\cal H}(\lambda)}}.
\label{free_derivative}
\end{equation}
Combining equations \eref{crossover_H} and \eref{free_derivative} the free energy difference can be expressed as
\begin{equation}
{\mathrm F}(1)-{\mathrm F}(0)=\int_0^1 d\lambda \frac{\partial {\mathrm F}(\lambda)}{\partial\lambda}=\int_0^1 d\lambda \<{\cal H}_2-{\cal H}_1\>_\lambda,
\label{free_diff}
\end{equation}
where $\<{\cal H}_2-{\cal H}_1\>_\lambda$ is the thermal average of the observable ${\cal H}_2-{\cal H}_1$ in the Gibbs ensemble of the crossover Hamiltonian ${\cal H}(\lambda)$ defined in \eeref{crossover_H}. For every $\lambda$ this average is accessible to standard MC simulations. Finally, the integral appearing in \eeref{free_diff} is performed numerically, yielding the free energy difference between the systems governed by the Hamiltonians ${\cal H}_2$ and ${\cal H}_1$, respectively. We apply \eeref{free_diff} with ${\cal H}_1$ as the Hamiltonian of the lattice $L\times L_\parallel\times L_\parallel$ with $(cs,+)$ or $(cs,o)$ {\bc} and with ${\cal H}_2$ as the Hamiltonian of the lattice $(L-1)\times L_\parallel\times L_\parallel$ plus a completely decoupled two-dimensional layer of non-interacting spins governed by the reduced Hamiltonian in \eeref{bc} with $\beta=0$, so that both Hamiltonians have the same configuration space\footnote{Here we are implicitly normalizing the free energy such that it vanishes for $\beta=0$. This choice amounts to a shift of the free energy which does not contribute to the critical Casimir force.}. Along these lines we have computed the free energy difference per area $\Delta{\cal F}$. A more detailed discussion of the implementation of this method can be found in \cite{PTD-10}.

\Tabletwo{\label{fitcs}Fit of MC data at criticality for $(cs,+)$ {\bc} to \eeref{fssansatz_crit}. $L_{\rm min}$ is the minimum lattice sizes used in the fit. $DOF$ denotes the number of degrees of freedom.}
\br
$L_{\rm min}$ & $f_{\rm bulk}(0)$ & $\Theta(\rho=0)$ & $E$ & $c$ & $\chi^2/DOF$\\
\hline
$8$ & $-0.0757368(1)$ & $2.398(5)$ & $-2.02(3)$ & $0.943(6)$ & $17.3/21$ \\
$12$ & $-0.0757368(2)$ & $2.406(11)$ & $-2.09(4)$ & $0.95(2)$ & $8.8/16$ \\
$16$ & $-0.0757367(4)$ & $2.41(2)$ & $-2.08(6)$ & $0.96(5)$ & $5.0/11$ \\
\br
\endTabletwo
\Tabletwo{\label{fitcsopen}Same as \tref{fitcs} for $(cs,o$) {\bc}.}
\br
$L_{\rm min}$ & $f_{\rm bulk}(0)$ & $\Theta(\rho=0)$ & $E$ & $c$ & $\chi^2/DOF$\\
\hline
$8$ & $0.07573692(3)$ & $0.494(3)$ & $-1.267(12)$ & $1.38(2)$ & $40.1/25$ \\
$12$ & $0.07573702(4)$ & $0.483(4)$ & $-1.28(2)$ & $1.22(4)$ & $13.2/20$ \\
$16$ & $0.07573697(5)$ & $0.492(7)$ & $-1.29(3)$ & $1.37(8)$ & $9.0/15$ \\
\br
\endTabletwo
At bulk criticality, \eeref{deltaF_full_exp} reduces to (see \eeref{theta_Theta})
\begin{eqnarray}
\Delta{\cal F}(t=0,L,L_\parallel) = f_{\rm bulk}(0)\nonumber \\
-\frac{1}{(L-1/2+c)^3}\left(\Theta\left(\rho=0\right)+ \frac{L-1/2+c}{L_\parallel} E\right)
\label{fssansatz_crit}
\end{eqnarray}
where analogous to \eeref{theta_Theta} we have defined $E\equiv E(\tau=0)$. Note that due to the choice of calculating the force at the intermediate thickness $L-1/2$ and due to corrections to scaling, the aspect ratio enters in \eeref{fssansatz_crit} as an effective aspect ratio $\tilde{\rho}=(L-1/2+c)/L_\parallel=\rho + O(1/L)$. In line with the discussion in \sref{sec:fss}, in \eeref{fssansatz_crit} we have expanded the critical Casimir force into powers of the aspect ratio $\rho$ up to linear order. It is interesting to observe that an eventual correction to \eeref{fssansatz_crit} $\propto\rho^2$ vanishes exactly (see the discussion in section 2.4 of \cite{PTD-10}).

In a series of MC simulations we have evaluated $\Delta{\cal F}(t=0,L,L_\parallel)$ for the {\bc} shown in \fref{cs}, setting $\beta=0.387721735$ and $D=0.655$; these values correspond to the bulk critical point of the model \cite{Hasenbusch-10}. We have sampled the lattice sizes $L=8$, $12$, $16$, $24$, $32$ and for each film thickness $L$ we have sampled the aspect ratios $\rho=L/L_\parallel=1/6$, $1/8$, $1/10$, $1/12$, $1/16$. We have fitted our MC results for $\Delta{\cal F}(t=0,L,L_\parallel)$ to \eeref{fssansatz_crit}, leaving $f_{\rm bulk}(0)$, $\Theta(\rho=0)$, $E$, and $c$ as free parameters. In \tref{fitcs} we report the fit results, for three minimum lattice sizes $L_{\rm min}$ taken into account for the fit. Inspection of the fit results reveals a good $\chi^2/DOF$ ratio ($DOF$ denotes the number of degrees of freedom in the fit). Moreover the fitted value of $f_{\rm bulk}(0)$ matches with the value $f_{\rm bulk}(0)=-0.0757368(4)$ reported in \cite{Hasenbusch-10}. There is also agreement with the values of $f_{\rm bulk}(0)$ obtained in \cite{PTTD-13} for the various {\bc} considered therein. The fitted values of $\Theta(\rho=0)$ and $E$ are in agreement with the previous estimates in \cite{PTD-10} which investigated $(cs,+)$ {\bc}: $\Theta=2.386(5)$, $E=-2.04(3)$. From the fits in \tref{fitcs} we extract $c=0.95(2)$. The knowledge of this parameter is instrumental in extracting the scaling function $E(\tau)$ from the MC data at a finite value of $L$ (see \sref{sec:mc:cs}). Unlike the universal quantities $\Theta(\rho=0)$ and $E$, the nonuniversal correction-to-scaling length $c$ is not expected to match the value found in \cite{PTD-10}. The reason for this is that here we have chosen $D=0.655$ whereas in \cite{PTD-10} the Blume-Capel model has been simulated for fixed $D=0.641$, which corresponds to an older determination of the coupling $D$ for which the model is improved, obtained as $D=0.641(8)$ in \cite{Hasenbusch-01}. Nevertheless, within the available precision the value $c=0.934(5)$ as determined in \cite{PTD-10} agrees with the present value $c=0.95(2)$\footnote{Note that, due to a different convention, the value of $c'=-0.066(5)$ reported in equation (84) of \cite{PTD-10} is related to $c$ via $c=c'+1$.}. The determination of $E$ allows one to test the validity of \eeref{thetasp} at criticality. In \cite{PTTD-13} we have studied the critical Casimir force for $(s,+)$ {\bc} (see \fref{bcstripes}), limited to the case $\varsigma=S_+/S_-=1$. In \cite{PTTD-13}, using the result $E=-2.04(3)$ of \cite{PTD-10}, we have found that \eeref{thetasp} holds rather well for $\varsigma=1$ and $\kappa=S_+/L\gtrsim 1$.

In order to study the critical Casimir amplitude for $(cs,o)$ {\bc}, we have computed $\Delta{\cal F}(t=0,L,L_\parallel)$ for lattice sizes $L=8$, $12$, $16$, $24$, $32$, $48$. We have sampled aspect ratios $\rho=L/L_\parallel=1/6$, $1/8$, $1/10$, $1/12$ for each lattice size, and $\rho=1/16$ for $L\le 32$. As for $(cs,+)$ {\bc} we have fitted $\Delta{\cal F}(t=0,L,L_\parallel)$ to \eeref{fssansatz_crit}, leaving $f_{\rm bulk}(0)$, $\Theta(\rho=0)$, $E$, and $c$ as free parameters. In \tref{fitcsopen} we report these fit results, for three minimum lattice sizes $L_{\rm min}$ taken into account for the fit. Fits with $L_{\rm min}=8$ lead to the somewhat large value of $\chi^2/DOF=1.6$, which may be due to residual scaling corrections. We note that the fitted values have a rather small error bar, which is due to the high statistical precision of the MC data: this could also contribute to an increase of the $\chi^2/DOF$ ratio because each data point gives an additive contribution to $\chi^2$ which is inversely proportional to the square of its error bar. As discussed in \sref{sec:fss:general}, $\Theta(\rho=0)$ is expected to attain the value $\Theta_{(+,o)}$ for laterally homogeneous $(+,o)$ {\bc}. Previous numerical investigations have reported $\Theta_{(+,o)}=0.497(3)$ \cite{Hasenbusch-11} and $\Theta_{(+,o)}=0.492(5)$ \cite{PTTD-13}. The values of $\Theta(\rho=0)$ fitted here (\tref{fitcsopen}) are in full agreement with the previous determinations, except for the fit results with $L_{\rm min}=12$, which is in marginal agreement with the values provided in \cite{Hasenbusch-11,PTTD-13}. A conservative judgement of the fit results in \tref{fitcsopen} leads to the final estimates
\begin{eqnarray}
&\Theta_{(cs,o)}(\rho=0) = \Theta_{(+,o)} = 0.492(7),\\
\label{Eopen}
&E_{(cs,o)} = -1.29(3),\\
&c = 1.36(10).
\end{eqnarray}

\begin{figure}[b]
\vspace{1em}
\includegraphics[clip=true,width=0.8\linewidth,keepaspectratio]{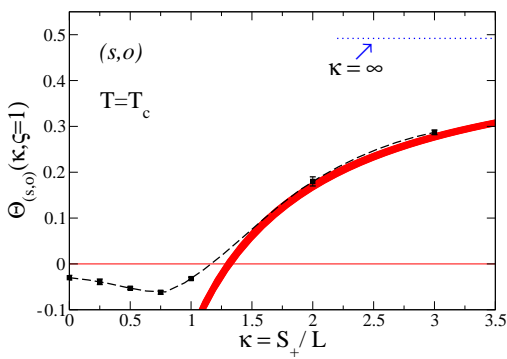}
\caption{Critical Casimir force amplitude $\Theta_{(s,o)}(\kappa, \varsigma=1)$ (full black squares) for $(s,o)$ {\bc} (see \fref{bcstripesopen}) as obtained in \cite{PTTD-13}. The dashed line provides a smooth interpolation. The thick red line shows the estimate of the rhs of \eeref{thetaso}, with $\Theta_{(+,o)}=\theta_{(+,o)}(\tau=0)=0.492(5)$ and $E = E(\tau=0)=-1.29(3)$; its thickness reflects the uncertainties of $\Theta_{(+,o)}(0)$ and $E(0)$. These lines saturate at $\Theta_{(s,o)}(\kappa\rightarrow\infty,\varsigma=1)=\Theta_{(+,o)}=0.492(5)$ \cite{PTTD-13}, which is indicated by the dotted line. An analogous comparison within MFT can be found in figure 24 in \cite{PTTD-13}. For $(s,+)$ {\bc}, figure 7 in \cite{PTTD-13} shows the analogous comparison between actual MC data and the corresponding analytic expression for the asymptotic behavior $\kappa\rightarrow\infty$ (equation (13) in \cite{PTTD-13}).}
\label{critopen}
\end{figure}

The determination of $E_{(cs,o)}$ allows one to test the validity of \eeref{thetaso} at criticality for $(cs,o)$ {\bc}. To this end, we use the critical Casimir amplitude for $(s,o)$ {\bc}, as computed in \cite{PTTD-13} for $\varsigma=1$ and several values of $\kappa$. Accordingly, the critical Casimir amplitudes, as obtained in \cite{PTTD-13}, are: $\Theta_{(s,o)}(\kappa=3,\varsigma=1)=0.287(5)$, $\Theta_{(s,o)}(\kappa=2,\varsigma=1)=0.18(1)$, $\Theta_{(s,o)}(\kappa=1,\varsigma=1)=-0.032(3)$, $\Theta_{(s,o)}(\kappa=3/4,\varsigma=1)=-0.062(4)$, $\Theta_{(s,o)}(\kappa=1/2,\varsigma=1)=-0.053(3)$, and $\Theta_{(s,o)}(\kappa=1/4,\varsigma=1)=-0.039(6)$. These results can be compared with the rhs of \eeref{thetaso} for $\tau=0$, setting $\varsigma=1$, $\theta_{(+,o)}(\tau=0)=\ \Theta_{(+,o)}=0.492(5)$ \cite{PTTD-13}, and, as determined above from \eeref{Eopen}, $E_{(cs,o)}=-1.29(3)$. The resulting asymptotic estimates are: $\Theta_{(s,o)}(\kappa=3,\varsigma=1)=0.277(7)$, $\Theta_{(s,o)}(\kappa=2,\varsigma=1)=0.170(9)$, $\Theta_{(s,o)}(\kappa=1,\varsigma=1)=-0.15(2)$, $\Theta_{(s,o)}(\kappa=3/4,\varsigma=1)=-0.37(2)$, $\Theta_{(s,o)}(\kappa=1/2,\varsigma=1)=-0.80(3)$, and $\Theta_{(s,o)}(\kappa=1/4,\varsigma=1)=-2.09(6)$. The estimate obtained from \eeref{thetaso} agrees well for $\kappa\ge 2$, while for $\kappa\le 1$ there are large deviations from the actual value of the critical Casimir amplitude $\Theta_{(s,o)}(\kappa,\varsigma=1)=\theta_{(s,o)}(\tau=0,\kappa,\varsigma=1)$. This is in line with the analysis in \cite{PTTD-13}, which shows that the critical Casimir force exhibits a qualitatively different behavior for $\kappa <2$ and $\kappa >2$. In \fref{critopen} we show a comparison of the critical Casimir amplitude as obtained in \cite{PTTD-13} with the asymptotic estimate given by \eeref{thetaso}. We find good agreement for $\kappa\ge 2$; a smooth interpolation of the available critical Casimir force amplitudes suggests that \eeref{thetaso} is reliable for $\kappa\gtrsim 1.5$.

\subsection {Universal scaling functions for chemical steps}
\label{sec:mc:cs}
The determination of the critical Casimir force off criticality has been performed using the algorithm introduced in \cite{Hucht-07} and employed in \cite{Hasenbusch-09b,Hasenbusch-09c,Hasenbusch-09d,Hasenbusch-10c,Hasenbusch-11,PTTD-13,PT-13}. The method consists of computing of the free-energy density via a numerical integration over $\beta$. For the Hamiltonian in \eeref{bc}, the free-energy density ${\cal F}(t,L,L_\parallel)$ can be expressed as
\begin{equation}
{\cal F}(t,L,L_\parallel)= -\int_0^\beta d\beta' E(\beta',L,L_\parallel),
\label{F_from_E}
\end{equation}
where the reduced energy density $E(\beta',L,L_\parallel)$, defined as
\begin{equation}
\label{Edef}
E(\beta,L,L_\parallel) \equiv \frac{1}{V}\Bigg\<\sum_{<i j>}S_i S_j\Bigg\>,
\end{equation}
can be sampled by standard MC simulations. A subsequent numerical integration of \eeref{Edef} renders ${\cal F}(t,L,L_\parallel)$. Finally, by repeating this procedure for two film thicknesses $L$ and $L-1$, we can compute the free energy difference $\Delta{\cal F}(t,L,L_\parallel)$ as defined in \eeref{deltaF}. Although this method requires to sample $E(\beta',L,L_\parallel)$ for many values of $\beta'$, it is more beneficial than the coupling parameter approach (see previous subsection) in determining the full scaling function. In fact, a suitable numerical quadrature, such as Simpson's rule, allows one to use the same sampled reduced energies as integrations points in the computation of the integral in \eeref{F_from_E} for several values of the upper integration limit $\beta$. In the actual implementation of the method it is rather useful to introduce a lower nonzero cutoff for the integral appearing in \eeref{F_from_E}. This is the case because the critical Casimir force is active only in a narrow interval of temperatures around the critical point, so that $\Delta{\cal F}(t,L,L_\parallel)\simeq f_{\rm bulk}(t)$, for $\xi(t)\ll L$. In practice by computing the integral in \eeref{F_from_E} with a lower cutoff $\beta_0>0$, one obtains $\Delta{\cal F}(t,L,L_\parallel)-\Delta{\cal F}(t_0,L,L_\parallel)$, with $t_0=\beta_c/\beta_0-1$. Subsequently, $\Delta{\cal F}(t_0,L,L_\parallel)$ can be conveniently computed using the coupling parameter approach as in \sref{sec:mc:tc}, and has then to be added to the previous result. A detailed description of the implementation of the method can be found in \cite{PTTD-13}.

Using the method outlined above, in a series of MC simulations we have computed $\Delta{\cal F}(t,L,L_\parallel)$ for $L=8$, $12$, $16$, $24$, for aspect ratios $\rho=L/L_\parallel=1/6$, $1/8$, $1/10$, $1/12$, $1/16$, and for a set of temperatures close to the critical point. In order to extract the chemical step contributions both for $(cs,+)$ and for $(cs,o)$ {\bc}, we make use of the fact that \eeref{deltaF_full} can be written as
\begin{equation}
\Delta{\cal F}(t,L,L_\parallel) = A(t,L) + \frac{1}{L_\parallel}B(t,L),
\label{offtcansatz}
\end{equation}
with
\begin{eqnarray}
A(t,L) = &f_{\rm bulk}(t) \nonumber \\&- \frac{\theta\left(t\left((L-1/2+c)/\xi_0^+\right)^{1/\nu},\rho=0\right)}{(L-1/2+c)^3}, \nonumber \\
B(t,L) = &- \frac{E\left(t\left((L-1/2+c)/\xi_0^+\right)^{1/\nu}\right)}{(L-1/2+c)^2}.
\label{offAB}
\end{eqnarray}
For each value of $t$ and $L$, we have fitted $\Delta{\cal F}(t,L,L_\parallel)$ to \eeref{offtcansatz}, leaving $A(t,L)$ and $B(t,L)$ as free parameters. Finally, the scaling functions $\theta(\tau,\rho=0)$ and $E(\tau)$ are obtained by inverting \eeref{offAB}. To this end, we use the value of $c$, as extracted from the fits at criticality in \sref{sec:mc:tc}. The determination of $\theta(\tau,\rho=0)$ requires the subtraction of the bulk free energy density $f_{\rm bulk}(t)$. In \cite{PTTD-13} we have computed $f_{\rm bulk}(t)$ for an interval of temperatures around the critical point, achieving a precision of $10^{-8}$. On the other hand, the scaling function $E(\tau)$, which describes the chemical steps contribution to the critical Casimir force, does not require the knowledge of $f_{\rm bulk}(t)$; this is because the thermodynamic limit of the free energy does not depend on the shape and the {\bc} of the system, so that $f_{\rm bulk}(t)$ contributes to $A(t,L)$ but not to $B(t,L)$ which implies that $E(\tau)$ is determined by $B(t,L)$ only. In order to calculate the scaling variable $\tau$, one needs the values of the nonuniversal amplitude $\xi_0^+$ of the correlation length $\xi$ and of the critical exponent $\nu$. From \cite{Hasenbusch-10c} we infer $\xi_0^+=0.4145(4)$. As for the critical exponent $\nu$, we use the recent result $\nu=0.63002(10)$ from \cite{Hasenbusch-10}.

\begin{figure}[b]
\vspace{1em}
\includegraphics[clip=true,width=0.8\linewidth,keepaspectratio]{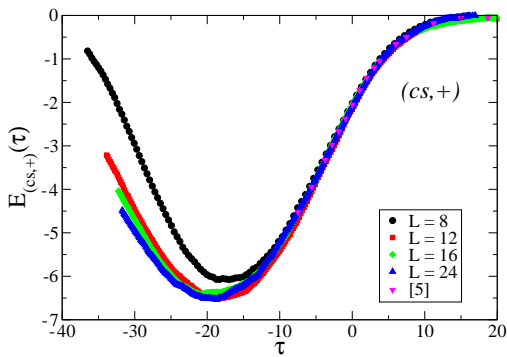}
\vspace{1em}
\caption{Universal scaling function $E(\tau)$ for $(cs,+)$ {\bc} which describes the chemical step contribution to the critical Casimir force. Scaling corrections have been suppressed by using $c=0.95(2)$. The error bars are a sum of the statistical error bars and of the uncertainty in $c$; the latter is the dominant contribution. The resulting error bars are of the size of the data points and are omitted. We compare our data with the previous results obtained in \cite{PTD-10} for the slab thickness $L=12$.}
\label{E}
\end{figure}

In \fref{E} we show the resulting universal scaling function $E(\tau)$ for $(cs,+)$ {\bc}, obtained by using $c=0.95(2)$. We observe a good scaling collapse for $\tau \gtrsim -14$. For $\tau \lesssim -14$, the data for $L=8$ systematically deviate from the other lattice sizes, signalling the presence of residual scaling corrections. For $L \ge 12$, there is only a small drift of the curves upon increasing $L$. In particular, the curves for $L=16$ and $L=24$ agree within the error bars. Accordingly, we can safely conclude that the curve for $L=24$ effectively realizes the FSS limit. We note that, aside from residual scaling corrections due to subleading irrelevant operators, for the large interval in $\tau$ displayed in \fref{E} additional corrections to scaling may originate from the so-called nonlinear scaling fields \cite{AF-83}, according to which the scaling field $t$ is replaced by an expansion $t+at^2+O(t^3)$. Such corrections may, at least partially, account for the deviation of the curve for $L=8$ from those with larger $L$. The universal scaling function $E(\tau)$ is always negative, implying that the chemical steps give rise to an attractive contribution to the critical Casimir force. It displays a minimum in the low-temperature phase at $\tau\simeq -20$; for $|\tau|\rightarrow \infty$ $E(\tau)$ vanishes. In fact, for $\tau\rightarrow -\infty$, the system approaches the ground state configuration, in which all spins take the same value as the upper surface (i.e., $S_i=+1$), thus forming an interface parallel to the lower surface. For such a configuration, a variation of the thickness $L$ by $1$ changes the film free energy by $f_{\rm bulk}$, which does not contribute to the derivative with respect to $L$ of the excess free energy, so that the critical Casimir force vanishes. This holds for any aspect ratio $\rho$. Therefore $E(\tau)$, which is the derivative of the Casimir force scaling function with respect to $\rho$ (see \eeref{thetaexp}), vanishes for $\tau\rightarrow -\infty$. In \fref{E} we also compare the present results with the previous ones $E(\tau)$ presented in \cite{PTD-10}. We find full agreement within the narrower interval in $\tau$ which was studied in \cite{PTD-10}. In \cite{PTD-10} we have also confirmed that the scaling function $\theta_{(cs,+)}(\tau,\rho=0)$ coincides with the mean value $(\theta_{(+,+)}(\tau)+\theta_{(+,-)}(\tau))/2$ of scaling functions for laterally homogeneous walls (see the discussion in \sref{sec:fss:cs}).

\begin{figure}
\includegraphics[clip=true,width=0.8\linewidth,keepaspectratio]{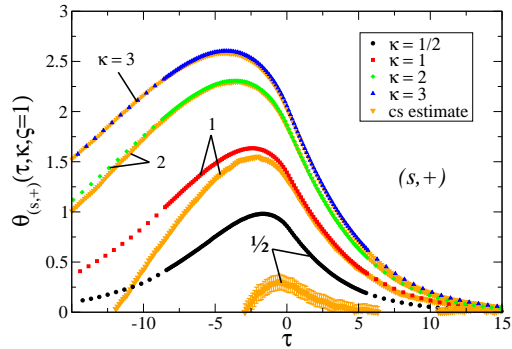}
\caption{Comparison between the universal scaling function $\theta_{(s,+)}(\tau,\kappa,\varsigma=1)$ for striped surfaces next to a homogeneous surface, for $\kappa=1/2$, $1$, $2$, and $3$ as obtained in \cite{PTTD-13}, and the chemical step estimate (cs estimate) given by the rhs of \eeref{thetasp}. The omitted statistical error bars are comparable with the symbol size.}
\label{comparison}
\end{figure}

The determination of $E_{(cs,+)}(\tau)$ allows one to test the reliability of \eeref{thetasp}. In \fref{comparison} we compare the universal scaling function $\theta_{(s,+)}(\tau,\kappa,\varsigma=1)$, which we have obtained in \cite{PTTD-13} for $\kappa=1/2$, $1$, $2$, and $3$, with the rhs of \eeref{thetasp}, which is computed by using the present results for $E_{(cs,+)}(\tau)$ and the results for $\theta_{(+,+)}(\tau)$ and $\theta_{(+,-)}$ provided by \cite{Hasenbusch-10c}. In line with a previous comparison carried out in \cite{PTTD-13}, we observe that for $\kappa\ge 2$ the chemical step estimate (i.e., the rhs of \eeref{thetasp}) reliably describes the scaling function $\theta_ {(s,+)}(\tau,\kappa,\varsigma=1)$. For $\kappa=1$, the rhs of \eeref{thetasp} agrees well with $\theta_ {(s,+)}(\tau,\kappa=1,\varsigma=1)$ for $\tau >0$, while there is a systematic deviation for $\tau<0$. For $\kappa\le 1/2$ the approximation for chemical stripes in terms of independent chemical steps breaks down.

\begin{figure}[b]
\vspace{1em}
\includegraphics[clip=true,width=0.8\linewidth,keepaspectratio]{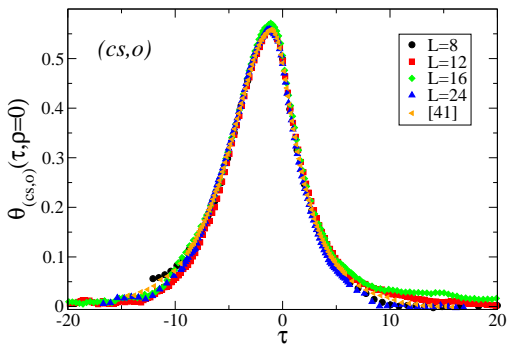}
\caption{Universal scaling function $\theta_{(cs,o)}(\tau,\rho=0)$ for $(cs,o)$ {\bc}. Scaling corrections have been suppressed by using $c=1.36(10)$. The error bars are a sum of the statistical error bars and of the uncertainty in $c$; the latter is the dominant contribution. Apart from the points for $\tau\lesssim -10$, the error bars are comparable with the size of the data points and are omitted. We compare our present results with previous ones for the scaling function $\theta_{(+,o)}(\tau)$ for $(+,o)$ {\bc} \cite{PTTD-13}, obtained for a slab width $L=24$. $\theta_{(cs,o)}(\tau,\rho=0)$ is expected to be identical with $\theta_{(+,o)}(\tau)$; the data fulfill this identity. An analogous comparison for $(cs,+)$ {\bc} is provided in figure 9 of \cite{PTD-10}.}
\label{thetao}
\end{figure}
\begin{figure}
\includegraphics[clip=true,width=0.8\linewidth,keepaspectratio]{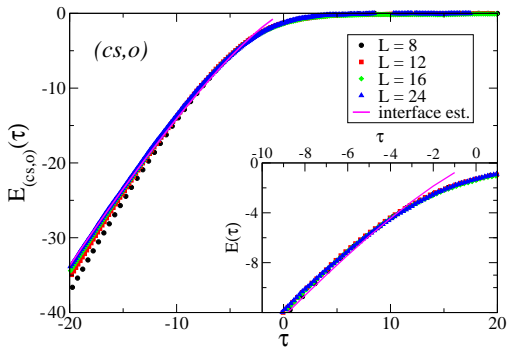}
\caption{Same as \fref{E} for $(cs,o)$ {\bc} and $c=1.36(10)$. We also compare our results with the interface estimate given by the rhs of \eeref{Ecsopen_interface}. The inset provides a magnification of the curves close to $\tau=0$. The error bars of the data are comparable with the symbol size and are omitted.}
\label{Eo}
\end{figure}

In \fref{thetao} we show the universal scaling function $\theta_{(cs,o)}(\tau,\rho=0)$ for $(cs,o)$ {\bc}, inferred from \eeref{offAB} by using $c=1.36(10)$ and $f_{\rm bulk}(t)$ as determined in \cite{PTTD-13}. We observe scaling collapse within the error bars. We also compare these results with previous ones for the scaling function $\theta_{(+,o)}(\tau)$ \cite{PTTD-13}. As expected from the discussion in \sref{sec:fss:cs}, $\theta_{(cs,o)}(\tau,\rho=0)$ coincides with the universal scaling function $\theta_{(+,o)}(\tau)$ for $(+,o)$ {\bc}. In \fref{Eo} we show the universal scaling function $E_{(cs,o)}(\tau)$, inferred from equations \eref{offtcansatz} and \eref{offAB} and by using $c=1.36(10)$. We observe a satisfactory scaling collapse with a small deviation, for strongly negative values of $\tau$, for the slab width $L=8$ only. In \fref{Eo} we also plot the rhs of \eeref{Ecsopen_interface}, which describes the behavior of $E_ {(cs,o)}(\tau)$ for $\tau\rightarrow -\infty$. To this end, we use the estimate of the universal amplitude ratio $R_\sigma=0.387(2)$ \cite{CHP-07}. The comparison of this curve with the data shows that the asymptotic behavior sets in for $\tau\lesssim -3$.

\begin{figure}
\vspace{1.5em}
\includegraphics[clip=true,width=0.8\linewidth,keepaspectratio]{comparisono}
\caption{Same as \fref{comparison} for $(s,o)$ {\bc}.}
\label{comparisono}
\end{figure}

As done for $(s,+)$ {\bc}, the knowledge of the scaling function $E_{(s,o)}(\tau)$ allows one to test the reliability of \eeref{thetaso}. In \fref{comparisono} we compare the universal scaling function $\theta_{(s,o)}(\tau,\kappa,\varsigma=1)$, obtained in \cite{PTTD-13} for $\kappa=1$ and $3$, with the rhs of \eeref{thetaso}, computed by using the present results for $E_{(cs,+)}(\tau)$ and the results for $\theta_{(+,o)}(\tau)$ in \cite{PTTD-13}. For $\kappa=3$ we find perfect agreement between the scaling function $\theta_{(s,o)}(\tau,\kappa=3,\varsigma=1)$ and the chemical step estimate given by \eeref{thetaso}. Consistent with the corresponding comparison at criticality shown in \fref{critopen}, the chemical step estimate displays a systematic deviation from $\theta_{(s,o)}(\tau,\kappa=1,\varsigma=1)$. The discrepancy increases upon decreasing $\tau$; in particular, the chemical step estimate does not capture the minimum of the critical Casimir force in the low-temperature phase. This finding is in agreement with the qualitative difference between the critical Casimir force for $\kappa<2$ and for $\kappa >2$ (see the corresponding discussion in \cite{PTTD-13}).

\section{Mean field theory}
\label{sec:mft}

In \cite{PTTD-13} a detailed comparison has revealed that the behavior of the scaling functions of the critical Casimir force for a chemically striped surface opposite to a homogeneous surface with $(+)$ or $(o)$ {\bc} is, on a \emph{qualitative} level, captured well by mean field theory (MFT). MFT provides the lowest order ($\varepsilon=0$) contribution to universal properties within an expansion in terms of $4-d=\varepsilon$. For the systems under consideration here, their qualitative features are consistent with the actual behavior in $d=3$.

The MFT order parameter profile $m\equiv u^{1/2}\langle\phi\rangle$ follows from minimizing the standard Landau-Ginzburg-Wilson fixed-point Hamiltonian \cite{Binder-83,Diehl-86}
\begin{eqnarray} 
  \label{eq:hamiltonian}
   \mathcal{H}[\phi]=&\int_V\,\upd^d r\,\left\{
        \frac{1}{2}(\nabla\phi)^2
       +\frac{\tilde\tau}{2}\phi^2
       +\frac{u}{4!}\phi^4
			 \right\} \nonumber\\
       &+\int_{\partial V}\upd^{(d-1)} r\left\{\frac{c(\vec{r})}{2}\phi^2-h_1(\vec{r})\phi\right\},
\end{eqnarray} 
where $\phi(\vec{r})$ is the spatially varying order parameter describing the critical medium, which completely fills the volume $V$ confined by the boundaries $\partial V$ in $d$-dimensional space; $\tilde\tau\propto t$ is proportional to the reduced temperature, and $u>0$ is the coupling constant. In the boundary term of the Hamiltonian, $c(\vec{r})$ is the surface enhancement, and $h_1(\vec{r})$ is an external surface field. In the strong adsorption limit, i.e., $(\pm)$ {\bc}, corresponding to the so-called normal surface universality class, the surface behavior is described by the renormalization-group fixed-point values $h_1\to\pm\infty$. Accordingly, the order parameter diverges close to the surface: $\phi|_{\partial V}\to\pm\infty$. The ordinary surface universality class, i.e., $(o)$ {\bc}, corresponds to the fixed point values $\{c=\infty,h_1=0\}$ and a vanishing order parameter $\phi|_{\partial V}=0$ \cite{Binder-83,Diehl-86}. 

Universal properties of the scaling functions of the critical Casimir force in $d=4$ can be determined within MFT up to logarithmic corrections and, generally, up to two independent nonuniversal amplitudes (such as the amplitude $B$ of the bulk order parameter $\langle\phi\rangle=\pm B|t|^\beta$ for $t<0$, where $\beta(d=4)=1/2$, and the amplitude $\xi_0^+$ of the correlation length). All quantities appearing in the bulk term of \eeref{eq:hamiltonian} can be expressed in terms of these amplitudes, i.e., $\tilde\tau=t (\xi_0^+)^{-2}$ and $u=6(B\xi_0^+)^{-2}$. We determine the critical Casimir force directly from the MFT order parameter profiles via the stress tensor \cite{Krech-97} up to an undetermined overall prefactor $\propto u^{-1}$. For this reason, the following MFT results are provided in terms of the critical Casimir amplitude $\Theta_{\pp}=8K^4(1/\sqrt{2}) (B\xi_0^+)^2 \simeq -47.2682 \times (B\xi_0^+)^2 =-283.6092/u$, where $K(k)$ is the complete elliptic integral of the first kind \cite{Krech-97}.

The spatially inhomogeneous MFT order parameter profile for the film geometry involving a chemical step is obtained via numerical minimization of $\mathcal{H}[\phi]$ using a quadratic finite element method. Analogous to the procedure described in detail in the previous sections, we first consider periodic {\bc} along the lateral directions, so that there is a chemically striped surface involving many chemical steps. The chemical step contributions $E_{(cs,+/o)}$ are subsequently determined by a least-square fit in the limit $\kappa\to\infty$. Numerically, we consider the case $S_+=S_-$ and $5<\kappa=S_+/L<15$.

The diverging order parameter profiles at those parts of the surface where there are $(+)$ or $(-)$ {\bc} are  implemented numerically via a short-distance expansion of the corresponding profile for the semi-infinite systems \cite{Binder-83,Diehl-86}, so that the MFT data presented below are subject to a numerical error which we estimate to be less than $3\%$ or $\pm0.05\times|\Theta_{\pp}|$ if the latter is bigger.

In \cite{PTTD-13}, for a chemically striped substrate next to a homogeneous substrate with $(+)$ or $(o)$ {\bc} we have determined in detail the asymptotic behavior $\propto\kappa^{-1}$ of the critical Casimir force scaling function for large $\kappa$ at $T=T_c$ and within MFT (see figures 23 and 24 in \cite{PTTD-13}). From the data in \cite{PTTD-13} one infers the MFT values $E_{(cs,+)}(\tau=0) \equiv E_{(cs,+)}\simeq-1.26 \times|\Theta_{(+,+)}|$ and $E_{(cs,o)}(\tau=0)  \equiv E_{(cs,o)}\simeq-0.43 \times|\Theta_{(+,+)}|$. From figures 23 and 24 of reference~\cite{PTTD-13} we also infer that equations \eref{thetasp}  and \eref{thetaso} are valid at $\tau=0$ already for values $\kappa\gtrsim1$. In the following, we extend these data by determining the scaling functions $E_{(cs,+/o)} (\tau)$ for a broad range of temperatures $\tau$.

\begin{figure}
\begin{center}
\includegraphics[clip=true,width=0.9\linewidth,keepaspectratio,clip=true]{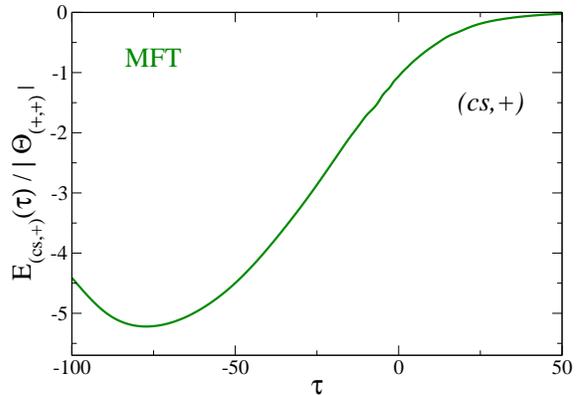}
\end{center}
\caption{Reduced universal scaling function $E_{(cs,+)}(\tau)$ for $(cs,+)$ {\bc} which corresponds to the chemical step contribution to the critical Casimir force, as obtained within MFT. The qualitative features of the MFT scaling function are similar to the ones obtained from the MC data in $d=3$ (compare \fref{E}). The numerical uncertainty increases for the most negative values of $\tau$. }
\label{eplus_mft}
\end{figure}
\begin{figure}
\begin{center}
\includegraphics[clip=true,width=0.9\linewidth,keepaspectratio,clip=true]{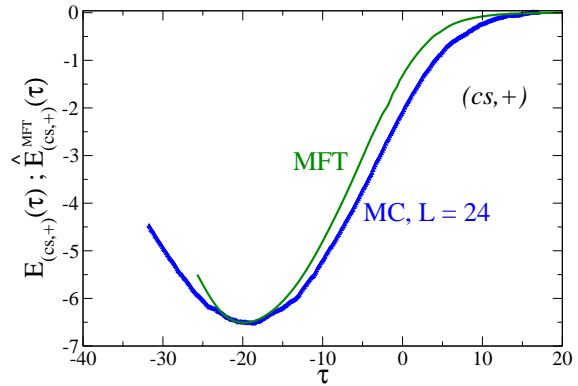}
\end{center}
\caption{Comparison between the rescaled universal scaling function $\hat{E}^{\rm MFT}_{(cs,+)}(\tau)$ for $(cs,+)$ {\bc} as obtained within MFT (full line, \eeref{eq:rescale}, $d=4$) and the MC simulation data (symbols, $d=3$). The MC data correspond to the ones for $L=24$ shown in \fref{E}. The MFT data have been rescaled such that both the depths and the positions of the minima of the two curves coincide.
}
\label{eplus_compare}
\end{figure}

In \fref{eplus_mft} we show the MFT universal scaling function $E_{(cs,+)}(\tau)$, which corresponds to the chemical step contribution to the critical Casimir force  for $(cs,+)$ {\bc}. The qualitative features of the MFT scaling function are similar to the ones for the scaling function in $d=3$ as obtained from MC data (see \fref{E}). In particular,  $E_{(cs,+)}(\tau)$ exhibits a minimum at $\tau<0$, and it reaches values of several multiples of $\Theta_{(+,+)}$. \Fref{eplus_compare} shows a comparison of the scaling functions for $d=3$ (MC) and $d=4$ (MFT).  In \fref{eplus_compare} the mean-field scaling function is rescaled linearly according to
\begin{equation}
  \label{eq:rescale}
  \hat{E}^{\rm MFT}_{(cs,+)}(\tau)\equiv
  \frac{E_{(cs,+)}(\tau_{\rm min})}{E_{(cs,+)}^{{\rm MFT}}(\tau_{\rm min}^{{\rm MFT}})}
 E^{\rm MFT}_{(cs,+)}\left(\frac{\tau_{{\rm min}}^{{\rm MFT}}}{\tau_{{\rm min}}}\tau\right)
\end{equation}
so that the position and the value of the minimum of the rescaled scaling function $\hat{E}^{{\rm MFT}}_{(cs,+)}$ agree with those of the MC data. In \eeref{eq:rescale} $\tau_{\rm min}$ and $\tau_{\rm min}^{\rm MFT}$ correspond to the position of the minimum of the scaling functions for $(cs,+)$ {\bc} in $d=3$ and $d=4$, respectively. From the MC data obtained for $L=24$ we infer the rough estimates $\tau_{\rm min}\simeq-19.9$ and $E_{(cs,+)}(\tau_{\rm min})\simeq-6.51$ in $d=3$ and $\tau^{\rm MFT}_{\rm min}\simeq -77.5$ and $E^{\rm MFT}_{(cs,+)}(\tau^{{\rm MFT}}_{\rm min})\simeq -5.22 \times|\Theta_{\pp}|$ in $d=4$. Note that the factor $|\Theta_{(+,+)}|$ drops out of $ \hat{E}^{\rm MFT}_{(cs,+)}(\tau)$.

\begin{figure}
\begin{center}
\includegraphics[clip=true,width=0.9\linewidth,keepaspectratio,clip=true]{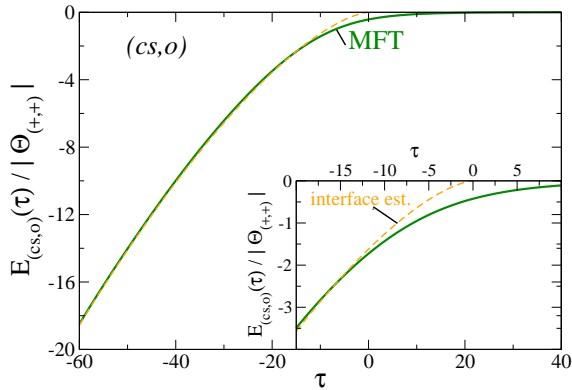}
\end{center}
\caption{Reduced universal scaling function $E_{(cs,o)}(\tau)$ for $(cs,o)$ {\bc} as obtained within MFT. The qualitative features of the MFT scaling function are similar to the ones obtained from the corresponding MC data (\fref{Eo}). Accordingly, also within MFT the interface estimate given by the rhs of \eeref{Ecsopen_interface} is reliable for $\tau\ll-1$.}
\label{eopen_mft}
\end{figure}

Similarly, the MFT universal scaling function  $E_{(cs,o)}(\tau)$ shown in \fref{eopen_mft} shares its qualitative features with those of the MC data (\fref{Eo}).  $E_{(cs,o)}(\tau)$ diverges for $\tau\to-\infty$ according to the interface estimate given in \eeref{Ecsopen_interface}, where within MFT $R_\sigma=\frac{2}{3}\sqrt{2}(B\xi_0^+)^2\simeq0.020 \times|\Theta_{\pp}|$ and $\mu=3/2$ (see also ref. \cite{PTTD-13}). On the other hand for $\tau\to+\infty$ the chemical step contribution vanishes.

Based on the results presented in detail in \cite{PTTD-13} we expect that the range of validity of \eeref{thetaso} within the geometrical parameter space spanned by $\kappa$ and $\varsigma$ is quantitatively similar for MFT and in $d=3$. Thus, based on equations \eref{thetasp} and \eref{thetaso} and for $\kappa\gg1$, the universal scaling functions $E_{(cs,+)}(\tau)$  and $E_{(cs,o)}(\tau)$, respectively, obtained within MFT may serve as building blocks for the calculation of critical Casimir forces emerging in more complex geometries.

\section{Summary}
\label{sec:summary}
We have studied the critical Casimir force for Ising film geometries $L\times L_\parallel\times L_\parallel$ with periodic lateral boundary conditions ({\bc}). The confining surface at the bottom is divided into two halves forming a \emph{c}hemical \emph{s}tep $(cs)$ with opposing adsorption preferences, while the upper confining surface features laterally homogeneous {\bc}, either with an adsorption preference $(+)$ or with open {\bc}, leading to $(cs,+)$ or $(cs,o)$ {\bc} for the film. These two types of {\bc} are illustrated in \fref{cs} and \fref{csopen}. Our main findings are as follows:
\begin{itemize}
  \item For small aspect ratios $\rho=L/L_\parallel$, the critical Casimir force varies linearly as a function of $\rho\rightarrow 0$. As discussed in \sref{sec:fss:cs} and in more detail in \cite{PTD-10}, this linear dependence is entirely due to the presence of the two individual chemical steps appearing in the geometries shown in figures \ref{cs} and \ref{csopen} with periodic lateral {\bc}. This dependence is characterized by a universal scaling function $E(\tau)$ (see equations (\ref{casimir_fss_leading}, \ref{Fcsplus}, \ref{Fcsopen}, \ref{forcecsrhoderivative}-\ref{forcecsrho})), where $\tau$ is the temperature-like scaling variable. From the MC data we have extracted the chemical step contribution $E_{(cs,+)}(\tau)$ and $E_{(cs,o)}(\tau)$ for $(cs,+)$ and $(cs,o)$ {\bc}, respectively, improving our previous results for $(cs,+)$ {\bc} \cite{PTD-10}.
  \item We have confirmed that in the limit $\rho\rightarrow 0$ the critical Casimir force for $(cs,o)$ {\bc} coincides with the force for laterally homogeneous $(+,o)$ {\bc} (see the discussion in \sref{sec:fss:cs}).
  \item As discussed in \sref{sec:fss:blocks}, the knowledge of the chemical step contribution $E(\tau)$ allows one to determine the critical Casimir force in the presence of a chemically striped substrate, in the limit of large stripe widths $S_+$ and $S_-$ (see figures \ref{bcstripes} and \ref{bcstripesopen} and equations \eref{thetasp} and \eref{thetaso}). We have compared this latter asymptotic expansion for large $\kappa\equiv S_+/L$ with the actual critical Casimir force for a chemically striped surface, which was determined in \cite{PTTD-13} for the case of alternating stripes of equal width $S_+=S_-$. In the case of $(cs,+)$ {\bc} the asymptotic behavior given by \eeref{thetasp} describes accurately the critical Casimir force for $\kappa\ge 1$, provided one is in the high-temperature phase $\tau >0$; for $\kappa \ge 2$ it is valid for the full critical temperature range. In the case of $(cs,o)$ {\bc} the asymptotic expansion given by \eeref{thetaso} taken at criticality $\tau=0$ agrees with the actual critical Casimir force amplitude for $\kappa\ge 2$. In the full temperature scaling range, however, the asymptotic expansion given by \eeref{thetaso} captures accurately the behavior of the critical Casimir force for $\kappa \ \gtrsim \  3$ only.
\item In \sref{sec:mft} we have presented the mean-field results ($d=4$) for both scaling functions $E_{(cs,+)}(\tau)$ and $E_{(cs,o)}(\tau)$, characterizing the chemical step contribution to the critical Casimir force. We find that the corresponding mean-field scaling functions (figures \ref{eplus_mft} and \ref{eopen_mft}) are qualitatively very similar to those obtained from MC simulation data in $d=3$, in particular after applying a suitable rescaling (\fref{eplus_compare}).
\end{itemize}
In conclusion, the determination of the chemical step contribution to the critical Casimir force acting on parallel substrates with various kinds of boundary conditions may allow for a straightforward approximate calculation of critical Casimir forces involving substrates with spatially complex chemical patterns. In this sense, even for curved surfaces next to chemically striped substrates equations \eref{thetasp} and \eref{thetaso} can be considered as a first-order improvement of the simple proximity force approximation, which neglects such line effects. This type of confinement is experimentally relevant for critical binary liquid mixtures which, as a solvent, lend themselves to experimental and theoretical studies of critical Casimir forces acting on solute particles in the vicinity of chemically striped substrates. They show  phenomena such as lateral forces \cite{SZHHB-08,TZGVHBD-11}, levitation \cite{TKGHD-10}, or self-assembly \cite{SZHHB-08,LLTHD-14} -- induced by critical fluctuations. So far, theoretical results for these setups have been obtained via the proximity force (or Derjaguin) approximation. Thus, the knowledge of the chemical step contribution may not only improve previous results but may also extend their range of validity in terms of a first-order extension of the Derjaguin approximation. We note that this concept has been extended also to geometrically structured substrates, by calculating within mean field theory a similar contribution to the critical Casimir force stemming from \emph{geometrical} steps at one of the confining surfaces \cite{THD-14}.

\end{document}